%% file: main.tex
\documentclass[conference]{IEEEtran}
\IEEEoverridecommandlockouts
\usepackage{cite}
\usepackage{amsmath,amssymb,amsfonts}
\usepackage{graphicx}
\usepackage{textcomp}
\usepackage{xcolor}
\def\BibTeX{{\rm B\kern-.05em{\sc i\kern-.025em b}\kern-.08em
    T\kern-.1667em\lower.7ex\hbox{E}\kern-.125emX}}
\input{newcommands}

\begin{document}

\title{Near Data Acceleration with \\
Concurrent Host Access
}

\author{\IEEEauthorblockN{Benjamin Y. Cho, Yongkee Kwon, Sangkug Lym, and Mattan Erez}
\IEEEauthorblockA{\textit{The University of Texas at Austin} \\
\{bjcho,yongkee.kwon,sklym,mattan.erez\}@utexas.edu}
}

\maketitle

\begin{abstract}
Near-data accelerators (NDAs) that are integrated with the main memory have the potential for significant power and performance benefits. Fully realizing these benefits requires the large available memory capacity to be shared between the host and NDAs in a way that permits both regular memory access by some applications and accelerating others with an NDA, avoids copying data, enables collaborative processing, and simultaneously offers high performance for both host and NDA. We identify and solve new challenges in this context: mitigating row-locality interference from host to NDAs, reducing read/write-turnaround overhead caused by fine-grain interleaving of host and NDA requests, architecting a memory layout that supports the locality required for NDAs and sophisticated address interleaving for host performance, and supporting both packetized and traditional memory interfaces.  We demonstrate our approach in a simulated system that consists of a multi-core CPU and NDA-enabled DDR4 memory modules. We show that our mechanisms enable effective and efficient concurrent access using a set of microbenchmarks, then demonstrate the potential of the system for the important stochastic variance-reduced gradient (SVRG) algorithm.
\end{abstract}


\input{tex/introduction}

\input{tex/background}

\input{tex/chonda2}
\input{tex/collaboration}
\input{tex/implementation}
\input{tex/methodology}
\input{tex/evaluation}
\input{tex/relatedwork}

\input{tex/conclusion}

\bibliographystyle{plain}
\bibliography{bib/ref}

\end{document}

%% file: newcommands.tex
\usepackage{graphicx}
\graphicspath{{fig/}}
\usepackage{xspace}
\usepackage{subfig}
\usepackage{xcolor}
\usepackage{amsmath}
\usepackage{array,multirow}
\usepackage{algorithm}
\usepackage[noend]{algpseudocode}
\usepackage{stfloats}
\usepackage[bottom,flushmargin]{footmisc}
\usepackage[colorinlistoftodos,prependcaption]{todonotes}
\usepackage{soul}

\newcommand{\mcut}[2][]{}
\newcommand{\bcut}[2][]{}

\newcommand{\scratch}[1]{}

\usepackage{makecell}                                                                  
\usepackage{xcolor, colortbl}   
\usepackage{tabu}  
\usepackage{array,booktabs,multirow}

\newcolumntype{C}{>{\centering\arraybackslash}p{8em}}
\DeclareMathSymbol{\minus}{\mathord}{operators}{"2D}

\definecolor{mygreen}{RGB}{0,90,0}

\newcommand{\fig}[1]{Figure~\ref{#1}\xspace}
\newcommand{\tab}[1]{Table~\ref{#1}\xspace}
\newcommand{\sect}[1]{Section~\ref{#1}\xspace}

\newcommand{\mattan}[1]{\textcolor{purple}{ME: #1\xspace}}

\newcommand{\meadd}[1]{\textcolor{black}{#1}}
\newcommand{\medel}[1]{\textcolor{purple}{}}
\newcommand{\ykadd}[1]{\textcolor{black}{#1}}

\newcommand{\mereplace}[2]{{\color{purple}{#1}{}}}

\newcommand{\ra}[1]{\renewcommand{\arraystretch}{#1}}

\newcommand{\hpcacut}[1]{}

\newcommand{\mymedskip}{\smallskip}

%% file: tex/introduction.tex

\section{Introduction} 
\label{sec:intro}

Processing data in or near memory using \emph{near data accelerators} (NDAs) is attractive for applications with low temporal locality and low arithmetic intensity. NDAs help by 
performing computation close to data, saving power and utilizing proximity to overcome the bandwidth bottleneck of a main memory ``bus'' (e.g.,~\cite{stone1970pim,kogge1994execube,gokhale1995processing,kogge1997processing,patterson1997case,kang1999flexram,guo20143d,farmahini2015nda,ahn2015pim,ahn2016scalable,asghari2016chameleon,gao2017tetris,alian2018nmp,alian2019netdimm,liu2018processing,boroumand2019conda}).
Despite decades of research and recent demonstration of true NDA technology~\cite{upmem,alian2018nmp,ibm_pim_dimm,pawlowski2011hybrid,nair2015active}, many challenges remain for making NDAs practical, especially in the context of \emph{main-memory NDA}.

In this paper we address several of these outstanding issues in the context of an NDA-enabled main memory. Our focus is on memory that can be concurrently accessed both as an NDA and as a memory. Such memory offers the powerful capability for the NDA and host processor to collaboratively process data  without costly data copies. Prior research in this context is limited to fine-grained NDA operations of, at most, cache-line granularity. However, we develop techniques for coarse-grain NDA operations that amortize host interactions across processing entire DRAM rows. 
At the same time, our NDA does not block host memory access, even when the memory devices are controlled directly by the host (e.g., a DDRx-like DIMM), which can reduce access latency and ease adoption.

\begin{figure}[b!ht]
\centering
	\includegraphics[width=0.35\textwidth]{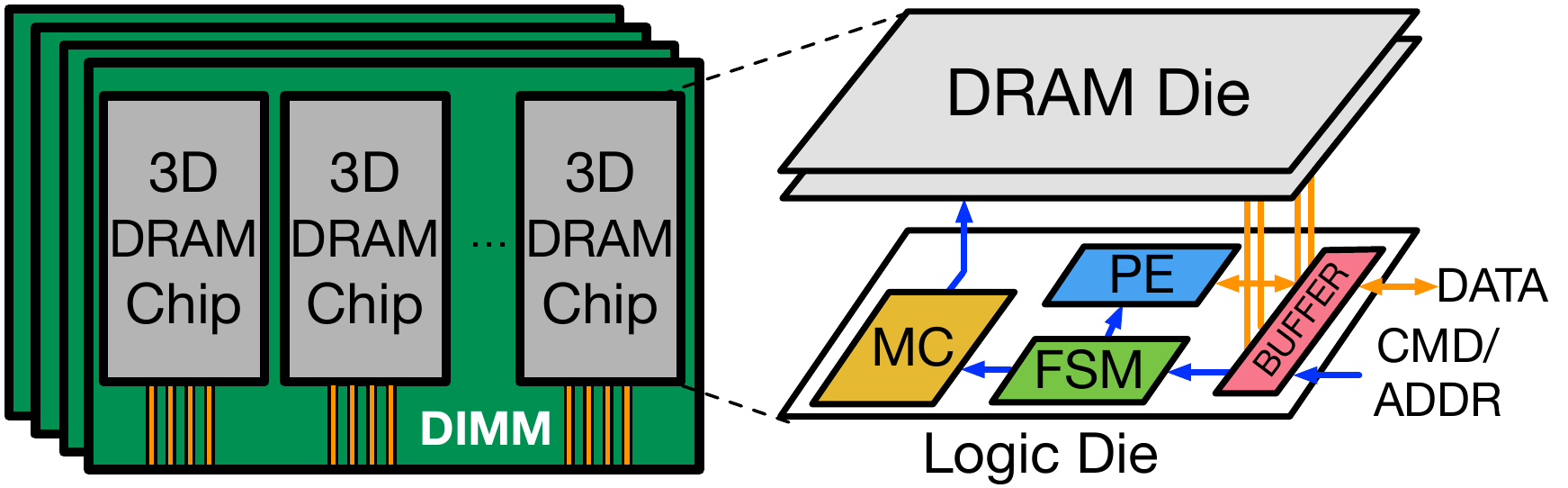}
	\caption{Exemplary NDA architecture.}
	\label{fig:baseline_nda}
	\vspace*{-4mm}
\end{figure}

\fig{fig:baseline_nda} illustrates an exemplary NDA architecture, which presents the challenges we address, and is similar to other recently-researched main-memory NDAs~\cite{farmahini2015nda,asghari2016chameleon,alian2018nmp}. We choose a DIMM-based memory system because it offers the high capacity required for a high-end server's main memory.
Each DIMM is composed of multiple chips, with one or more DRAM dice stacked on top of a logic die in each chip, using a low-cost commodity 3DS-like approach. Processing elements (PEs) and a memory controller are located on the logic die. Each PE can access memory internally through the NDA memory controller. These local NDA accesses must not conflict with external accesses from the host (e.g., a CPU). A rank that is being accessed by the host cannot at the same time serve NDA requests, though the bandwidth of all other ranks in the channel can be used by the NDAs. 
There is no communication between PEs 
other than through the host. While not identical, recent commercial NDA-enabled memories exhibit similar overall characteristics~\cite{upmem,ibm_pim_dimm}. 


\meadd{Surprisingly, no prior work on NDA-enabled main memory examines the architectural challenges of simultaneous and concurrent access to memory devices from both the host and NDAs. In this work, we address two key challenges for enabling performance-efficient NDAs in a memory system that supports concurrent access from both a high-performance host and the NDAs.}

The first challenge is that interleaved accesses may hurt memory performance because they can both decrease row-buffer locality and introduce additional read/write turnaround penalties. The second challenge is that each NDA can process kernels that consume entire arrays, though all the data that a single operation processes must be local to a PE (e.g., a memory chip). Therefore, enabling cooperative processing requires that host physical addresses are mapped to memory locations (channel, rank, bank, etc.) in a way that both achieves high host-access performance (through effective and complex interleaving) and maintains NDA locality across all elements of all operands of a kernel.
We note that these challenges exist when using either a packetized interface, where the memory-side controller interleaves accesses between NDAs and the host, or a traditional host-side memory controller that sends explicit low-level memory commands.  

\emph{For the first challenge (managing concurrent access)}, we identify reduced row-buffer locality because of interleaved host requests as interfering with NDA performance. In contrast, it is the increased read/write turnaround frequency resulting from NDA writes that mainly interfere with the host. We provide two solutions in this context. First, we develop a new bank-partitioning scheme that limits interference to just those memory regions that are shared by the host and NDAs, thus enabling colocating host-only tasks with tasks that use the NDAs. This new scheme is the first that is compatible with huge pages and also with the advanced memory interleaving functions used in recent processors. 
Partitioning mitigates interference from the host to the NDAs and substantially boosts their performance (by $1.5-2\times$). 

Second, 
we control interference on shared ranks by opportunistically issuing NDA memory commands to those ranks that are even briefly not used by the host and curb NDA to host interference with mechanisms that can throttle NDA requests, either selectively when we predict a conflict (\emph{next-rank prediction}) or stochastically.

\emph{For the second challenge (NDA operand locality)}, we enable fine-grain collaboration by architecting a new data layout that preserves locality of operands within the distributed NDAs while simultaneously affording parallel accesses by the high-performance host. This layout requires minor modifications to the memory controller and utilizes coarse-grain allocations and physical-frame coloring in OS memory allocation. This combination allows large arrays to be shuffled across memory devices (and their associated NDAs) in a coordinated manner such that they remain aligned in each NDA. This is crucial for coarse-grain NDA operations that can achieve higher performance and efficiency than cacheline-oriented fine-grain NDAs (e.g.,~\cite{ahn2015pim,kim2017toward,hsieh2016transparent}). 


\emph{An additional and important challenge} exists in systems where the host maximizes its memory performance by directly controlling memory devices \meadd{rather than relying on a packetized interface~\cite{pawlowski2011hybrid,hadidi2018performance}}. Adding NDA capabilities requires providing local memory controllers near memory in addition to the host ones\meadd{, which introduces a coordination challenge}. We coordinate memory controllers and ensure a consistent view of bank and timing state \meadd{with only minimal signaling that does not impact performance by  replicating the controller finite state machines (FSMs) at both the NDA and host sides of the memory channels}.
Replicating the FSM requires all NDA accesses to be determined only by the NDA operation (known to the host controller) and any host memory operations. Thus, no explicit signaling is required from the NDAs back to the host. We therefore require that for non-packetized NDAs, each NDA operation has a deterministic access pattern for all its operands (which may be arbitrarily fine-grained). 

In this paper, we introduce \emph{Chopim}, a SW/HW holistic solution that enables concurrent host and NDA access to main memory by addressing the challenges above with fine temporal access interleaving to physically-shared memory devices. We perform a detailed evaluation both when the host and NDA tasks process different data and when they collaborate on a single application. We demonstrate that Chopim enables high NDA memory throughput (up to 97\% of unutilized bandwidth) while maintaining host performance. Performance and scalability are better than with prior approaches of partitioning ranks and only allowing coarse-grain temporal interleaving, or with only fine-grain NDA operations. 

We demonstrate the potential of host and NDA collaboration by studying a machine-learning application (logistic regression with stochastic variance-reduced gradient descent~\cite{johnson2013accelerating}). We map this application to the host and NDAs such that the host stochastically updates weights in a tight inner loop that utilizes the speculation and locality mechanisms of the CPU while NDAs concurrently compute a correction term across the entire input data that helps the algorithm converge faster. Collaborative and parallel NDA and host execution can speed up this application by $2\times$ compared to host-only execution and $1.6\times$ compared to non-concurrent host and NDA execution. We then evaluate the impact of colocating such an accelerated application with host-only tasks.


In summary, we make the following main contributions:
\begin{itemize}
\item We identify new challenges in concurrent access to memory from the host and NDAs: bank conflicts from host accesses curb NDA performance and read/write-turnaround penalties from NDA writes lower host performance.
\item We reduce bank conflicts with a new bank partitioning architecture that, for the first time, is compatible with both huge pages and sophisticated memory  interleaving.
\item To decrease read/write-turnaround overheads, we throttle NDA writes with two mechanisms: \textit{next-rank prediction} delays NDA writes to the rank actively read by the CPU; and \textit{stochastic issue} throttles NDA writes randomly at a configurable rate.
 
\item We develop, also for the first time, a memory data layout that is compatible with both the host and NDAs, enabling them to collaboratively process the same data in parallel while maintaining high host performance with sophisticated memory address interleaving.

\item To show the potential of collaboratively processing the same data, we conduct a case study of an important ML algorithm that leverages the fast CPU for its main training loop and the high-BW NDAs for summarization steps that touch the entire dataset. We develop a variant that executes on the NDAs and CPU in parallel, which increases speedup to 2X.

\end{itemize}

%% file: tex/background.tex
\section{Background}
\label{sec:background}


\mymedskip
\noindent\textbf{\textit{DRAM Basics.}}
A memory system is composed of memory channels that operate independently. In each memory channel, one or more memory modules (DIMMs) share command/address (C/A) and data bus. A DIMM is usually composed of one or two physical ranks where all chips in the same rank operate together. Each chip and thus rank is composed of multiple banks and bank state is independent. Each bank can be in an opened or closed state and, if opened, which row is opened. To access a certain row, the target row must be opened first. If another row is already open, it must be closed before the target row is opened, which is called \textit{bank conflict} and increases access latency. The DRAM protocol specifies the timing parameters and protocol accessing DRAM. These are managed by a per-channel memory controller.

\mymedskip
\noindent\textbf{\textit{Address Mapping.}}
The memory controller translates OS-managed physical addresses into DRAM addresses, which are composed of indices to channel, rank, bank, row, and column. Typically, memory controllers follow the following policies in their address mapping to minimize access latency: interleaving address across channels with fine granularity is beneficial since they can be accessed independently from each other. On the other hand, ranks are interleaved at coarse granularity since switching to other ranks in the same channel incurs a penalty. In addition, XOR-based hash mapping functions are used when determining channel, rank, and bank addresses to maximally exploit bank-level parallelism. This also minimizes bank conflicts when multiple rows are accessed with the same access pattern since the hash function shuffles the bank address order \cite{zhang2000permutation}. To accomplish this, some row address bits are used along with channel, rank, and bank address bits~\cite{pessl2015reverse}.

\mymedskip
\noindent\textbf{\textit{Write-to-Read Turnaround Time.}}
In general, interleaving read and write DRAM transactions incurs higher latency than issuing the same transaction type back to back. Issuing a read transaction immediately following a write suffers from particularly high penalty. The memory controller issues the write command and loads data to the bus after tCWL cycles. Then, data is transferred for tBL cycles to the DRAM device and written to the cells. The next read command can only be issued after tWTR cycles, which guarantees no conflict on the IO circuits in DRAM. The high penalty stems from the fact that the actual write happens at the end of the transaction whereas a read happens right after it is issued. For this reason, the opposite order, read to write, has lower penalty. 

\mymedskip
\noindent\textbf{\textit{NDA Basics.}}
Near-data accelerators add processing elements near memory to overcome the physical constraints that limit host memory bandwidth. \medel{Since memory channels are independent,} Host peak memory bandwidth is determined by the number of channels and peak bandwidth per channel. \meadd{Any NDA accesses on the memory side of a channel can potentially increase overall system bandwidth.} 
\meadd{For example. a memory module with multiple ranks offers more bandwidth in the module than available at the channel. Similarly, multiple banks on a DRAM die can also offer more bandwidth than available off of a DRAM chip.}
\medel{However, the number of ranks in the system does not affect the peak memory bandwidth of the host since only one rank per channel can transfer data to the host at any given time over the shared bus. On the other hand, near-data accelerators (NDAs) can access data internally without contending for the shared bus. This enables higher peak bandwidth than the host can achieve.}
However, because NDAs only offer a BW advantage when they access data in their local memory, data layout is crucial for performance. A naive layout may result in frequent data movement among NDAs and with the host. \medel{In this paper, we assume that inter-NDA communication is only done through the host (alternatives are discussed in~\cite{kim2013memory,poremba2017there}).} 

\mymedskip
\noindent\textbf{\textit{Baseline NDA Architecture.}}
Our work targets NDAs that are integrated within high-capacity memory modules such that their role as both main memory and as accelerators is balanced. Specifically, our baseline NDA devices are 3D-integrated within DRAM chips on a module (DIMM), similar to 3DS DDR4 \cite{ddr43ds} yet a logic die is added. DIMMs offer high capacity and predictable memory access\mcut{, which are the required features for main memory}. \meadd{Designs with similar characteristics include on-DIMM PEs~\cite{ibm_pim_dimm,alian2018nmp} and on-chip PEs within banks~\cite{upmem}.}
Alternatively, NDAs can utilize high-bandwidth devices, such as the hybrid memory cube (HMC) \cite{pawlowski2011hybrid} or high bandwidth memory (HBM) \cite{standard2013high}. These offer high internal bandwidth but have limited capacity and high cost due to numerous point-to-point connections to memory controllers \cite{asghari2016chameleon}. HMC provides capacity scaling via a network but this results in high access latency and cost. HBM does not provide such solutions. As a result, HBM devices are better for standalone accelerators than for main memory. 

\hpcacut{
\mattan{I copied this to intro. Probably need to rearrange some things here.}
\fig{fig:baseline_nda} illustrates our baseline NDA architecture. Each DIMM is composed of multiple chips, with one or more DRAM dice stacked on top of a logic die in each chip, using the low-cost commodity 3DS approach. Processing elements (PEs) and a memory controller are located on the logic die. Each PE can access memory internally through the NDA memory controller. However, this internal access cannot conflict with external accesses from the host CPU (host). Therefore, each rank is in either host or NDA access mode and only one can access it at any given time. The host uses chip-address memory-mapped registers to control the NDAs~\cite{farmahini2015nda}. 
}

\begin{table}\centering
  \ra{1.2}
  \small
\begin{tabular}{@{}llll@{}}\toprule
Operations & Description  & Operations & Description \\
\midrule
AXPBY & ${\vec{z} = \alpha \vec{x} + \beta \vec{y}}$ & DOT & ${c = \vec{x} \cdot \vec{y}}$ \\
AXPBYPCZ & ${\vec{w} = \alpha \vec{x} + \beta \vec{y} + \gamma \vec{z}}$ & 	NRM2 & ${c = \sqrt{\vec{x} \cdot \vec{x}}}$ \\
AXPY & ${\vec{y} = \alpha \vec{y} + \vec{x}}$ & SCAL & ${\vec{x} = \alpha \vec{x}}$ \\
COPY & ${\vec{y} = \vec{x}}$ & GEMV & ${\vec{y} = A\vec{x}}$ \\
XMY & ${\vec{z} = \vec{x} \odot \vec{y}}$ &   &             \\
\bottomrule
\end{tabular}
\caption{Example NDA operations used in our case-study application. Chopim is not limited to these operations.}
\label{tab:nda_ops} 
\vspace*{-4mm}
\end{table}

\mymedskip
\noindent\textbf{\textit{Coherence.}}
Coherence mechanisms between the host and NDAs have been studied in prior NDA work~\cite{ahn2015pim,boroumand2019conda,boroumand2016lazypim} and can be used as is with Chopim. We therefore do not focus on coherence in this paper. In our experiments, we use the existing coherence approach of explicitly and infrequently copying the small amount of data that is not read-only using cache bypassing and memory fences.  

\mymedskip
\noindent\textbf{\textit{Address Translation.}}
\meadd{Application use of NDAs requires virtual to physical address translation. Some prior work~\cite{hsieh2016accelerating,hong2016accelerating,gao2015practical} proposes  address translation within NDAs to enable independent NDA execution without host assist. This increases both NDA and system complexity. As an alternative, NDA operations can be constrained to only access data within a physical memory region that is contiguous in the virtual address space. Hence, translation is performed by the host when targeting an NDA command at a certain physical address. This has been proposed for both very fine-grain NDA operations within single cache lines~\cite{ahn2015pim,ahn2016scalable,bssync,kim2017toward,nai2017graphpim} and NDA operations within a virtual memory page~\cite{oskin1998active}.}
\medel{Before the host and/or NDAs accesses memory, logical-to-physical address translation should be done. One possible approach is to make the host OS do the address translation for all host and NDA accesses. On the other hand, there are prior work \cite{hsieh2016accelerating,hong2016accelerating} that attempts to do address translation with NDAs to enable independent NDA execution without host's assist.} In this paper, \meadd{we use host-based translation because of its low complexity and only check bounds within the NDAs for protection.} \medel{choose the first approach where the host has direct control over NDAs.}

\mymedskip
\noindent\textbf{\textit{NDA Workloads.}}
We focus on NDA workloads for which the host inherently cannot outperform an NDA. These exhibit low temporal locality and low arithmetic intensity and are  bottlenecked by peak memory bandwidth. By offloading such operations to the NDA, we mitigate the bandwidth bottleneck by leveraging internal memory module bandwidth. Moreover, these workloads typically require simple logic for computation and integrating such logic within DRAM chips/modules is practical because of the low area and power overhead. 


Fundamental linear algebra matrix and vector operations satisfy these criteria. Dense \meadd{vector and matrix-vector operations, which are prevalent in machine learning primitives,} are particularly good candidates because of their deterministic and regular memory access patterns and low arithmetic-intensity.
For example, prior work off-loads matrix and vector operations of deep learning workloads to utilize high near-memory BW~\cite{kim2016neurocube,gao2017tetris}.
Also, Kwon et al. propose to perform element-wise vector reduction operations needed for a deep-learning-based recommendation system to NDAs~\cite{kwon2019tensordimm}.
In this paper, we focus on accelerating the dense matrix and vector operations summarized in \tab{tab:nda_ops}. We demonstrate and evaluate their use in the SVRG application in Section \ref{sec:collaboration}. \meadd{Note that we use these as a concrete example, but our contributions generalize to other NDA operations.}


NDA execution of graph processing has also been proposed because graph processing can be bottlenecked by peak memory bandwidth because of low temporal and spatial locality~\cite{nai2017graphpim,zhang2018graphp,song2018graphr,ahn2016scalable,ahn2015pim}. We do not consider graph processing in this paper because we do not innovate in this context.

%% file: tex/chonda2.tex
\section{Chopim}
\label{sec:chonda}

We develop Chopim with four main connected goals that push the state of the art: (1) enable fine-grain interleaving of host and NDA memory requests to the same physical memory devices while mitigating the impact of their contention; (2) permit the use of coarse-grain NDA operations that process long vector instructions/kernels; (3) simultaneously support the locality needed for NDAs and the sophisticated memory address interleaving required for high host performance; and (4) integrate with both a packetized interface and a traditional host-controlled DDRx interface.
We detail our solutions in this section after summarizing the need for a new approach.

\hpcacut{
Two Our  approach on managing concurrent access on different data is for NDAs to opportunistically utilize even a brief moment that the host does not access memory. To leverage these short idle periods, overhead for memory ownership switching should be minimized. In this way, the internal memory bandwidth can be fully utilized yet gives negligible impact on the host performance. In this section, we present our mechanisms that enable fine-grain interleaving between host and NDA accesses. Note that we can allocate more time and allow an exclusive access for NDA executions based on certain policy but we do not explore this in this paper. 

In addition, our approach on concurrent access to the single copy of shared data is to use memory controller's address mapping as is while localizing NDA operands at memory allocation and execution time. As no data reorganization is required between host and NDA access phases, this data layout enables concurrent and collaborative processing between the host and NDAs on the shared data. 
}

\begin{figure}[t!bh]
\centering
	\includegraphics[width=0.48\textwidth]{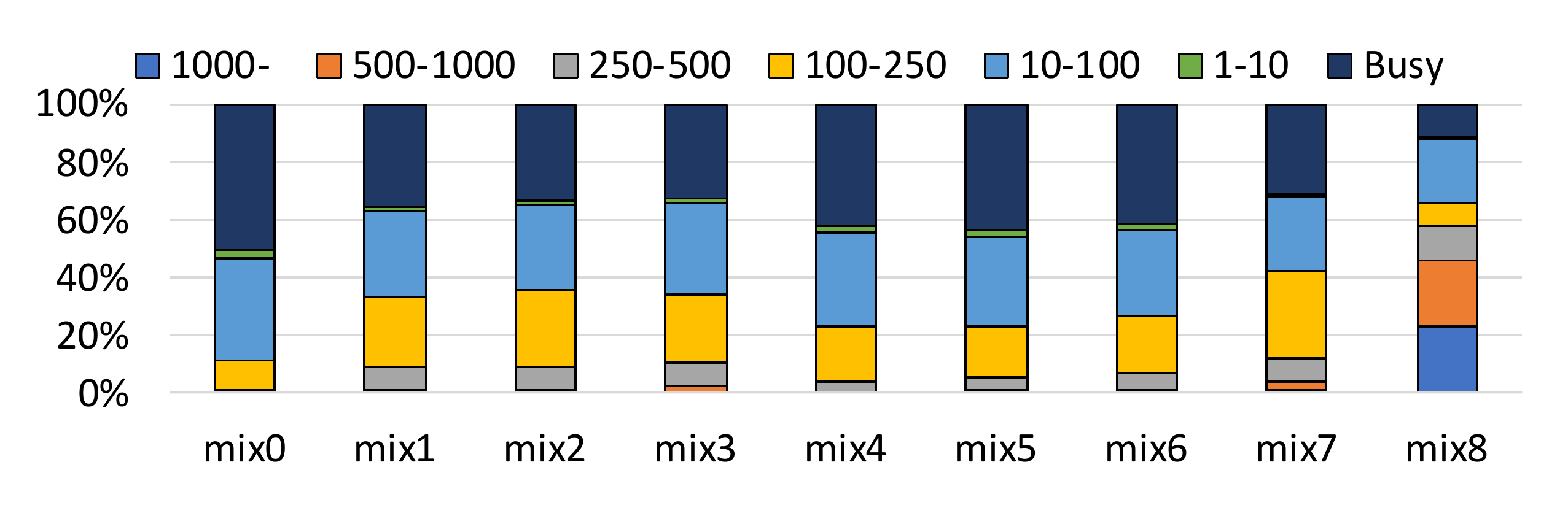}
	\vspace*{-2mm}
	\caption{Rank idle-time breakdown vs. idleness granularity.}
	\label{fig:motiv_rank_idle}
	\vspace*{-2mm}
\end{figure}

\medskip
\noindent\textbf{\textit{The need for fine-grain access interleaving with opportunistic NDA issue.}}
\label{subsec:nda_issue}
An ideal NDA  opportunistically issues NDA memory requests whenever a rank is idle from the perspective of the host. This is simple to do in a packetized interface where a memory-side controller schedules all accesses, but is a challenge in a traditional memory interface because the host- and NDA-side controllers must be synchronized. Prior work proposed dedicating some ranks to NDAs and some to the host or coarse-grain temporal interleaving~\cite{farmahini2015nda,asghari2016chameleon}. The former approach contradicts one of our goals as devices are not shared. The latter results in large performance overhead because it cannot effectively utilize periods where a rank is naturally idle due to the host access pattern. \fig{fig:motiv_rank_idle} shows that for a range of multi-core application mixes (methodology in \sect{sec:method}), the majority of idle periods are shorter than 100 cycles with the vast majority under 250 cycles.  \emph{Fine-grain access interleaving is therefore necessary. }
\vspace*{-2mm}

\medskip
\noindent\textbf{\textit{The need for coarse-grain NDA vector/kernel operations.}}
\label{subsec:launch_ovhd}
Fine-grain access interleaving is simple if each NDA command only addresses a single cache block region of memory. Such fine-grain NDA operations have indeed been discussed in prior work~\cite{ahn2015pim,ahn2016scalable,bssync,nai2017graphpim}. One overhead of this fine-grain approach is that of issuing numerous NDA commands, with each requiring a full memory transaction that occupies both the command and data channels to memory. Issuing NDA commands too frequently degrades host performance, while infrequent issue underutilizes the NDAs.
Coarse-grain NDA vector operations that operate on multiple cache blocks mitigate contention on the channel and improve overall performance. The vector width, ${N}$, is specified for each NDA instruction. As long as the operands are contiguous in the DRAM address space, one NDA instruction can process numerous data elements without occupying the channel. Coarse-grain NDA operations are therefore desirable, but \emph{introduce the data layout, memory contention, and host--NDA synchronization challenges which Chopim solves}.

\subsection{Localizing NDA Operands while Distributing Host Accesses}
\label{subsec:data_layout}
To execute the N-way NDA vector instructions, all the operands of each NDA instruction must be fully contained in a single rank \meadd{(single PE)}. If necessary, data is first copied from other ranks prior to launching an NDA instruction. If the reuse rate of the copied data is low, this copying overhead will dominate the NDA execution time and contention on the memory channel will increase due to the copy commands. 

\begin{figure}[t!]
\centering
        \includegraphics[width=0.52\textwidth]{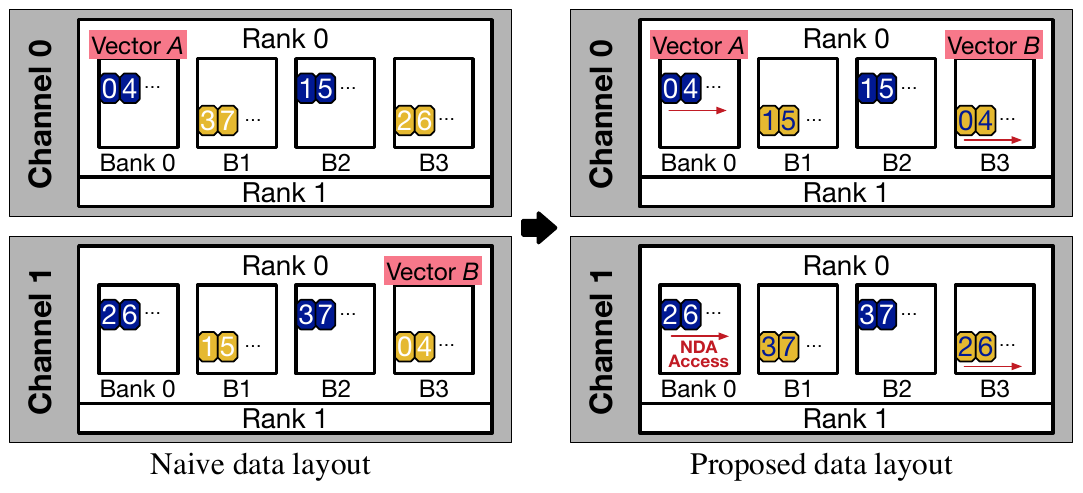}
	\caption{Example data layout across ranks for concurrent access of the COPY operation (B[i] = A[i]). With naive data layout (left), elements with the same index are located in different ranks. With our proposed mechanism (right), elements with the same index are co-located. NDAs access contiguous columns starting from the base of each vector.}
	\label{fig:rank_layout}
\end{figure}

\emph{We solve this problem} in Chopim by laying out data such that all the operands are localized to each NDA at memory allocation time. Thus, copies are not necessary. This is challenging, however, because the host memory controller uses complex address interleaving functions to maximally exploit channel, rank, and bank parallelism for arbitrary host access patterns. Hence, arrays that are contiguous in the host physical address space are not contiguous in physical memory and are shuffled across ranks.\medel{, possibly in a physical-address dependent manner.} 
This challenge is illustrated in the left side of \fig{fig:rank_layout}, where two operands of an NDA instructions are shuffled differently across ranks and banks. The layout resulting from our approach is shown at the right of the figure, where arrays (operands) are still shuffled, but both operands follow the same pattern and remain correctly aligned to NDAs without copy operations. Note that alignment is to rank because that corresponds to an NDA partition. 

\medskip
\noindent\textbf{\textit{Data layout across ranks.}} 

\begin{figure}[t!]
\centering
  \begin{minipage}[t]{0.4\textwidth}
			\subfloat [Baseline (Skylake~\cite{pessl2016drama})] {
				\includegraphics[width=\textwidth]{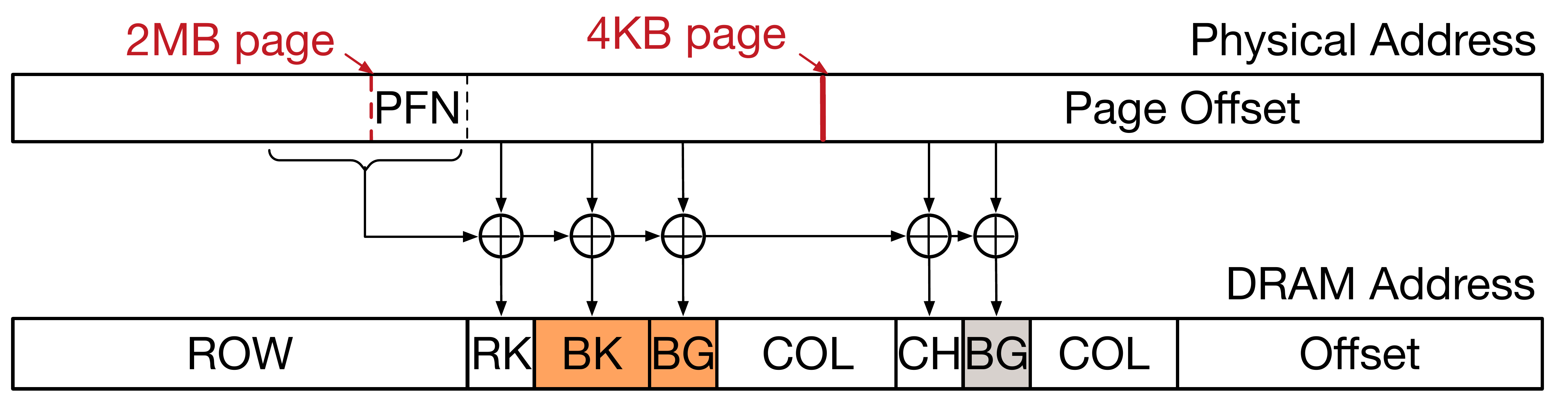}
				\label{fig:baseline_addr_map}
			} \\
			\subfloat [Proposed (for bank partitioning)] {
				\includegraphics[width=\textwidth]{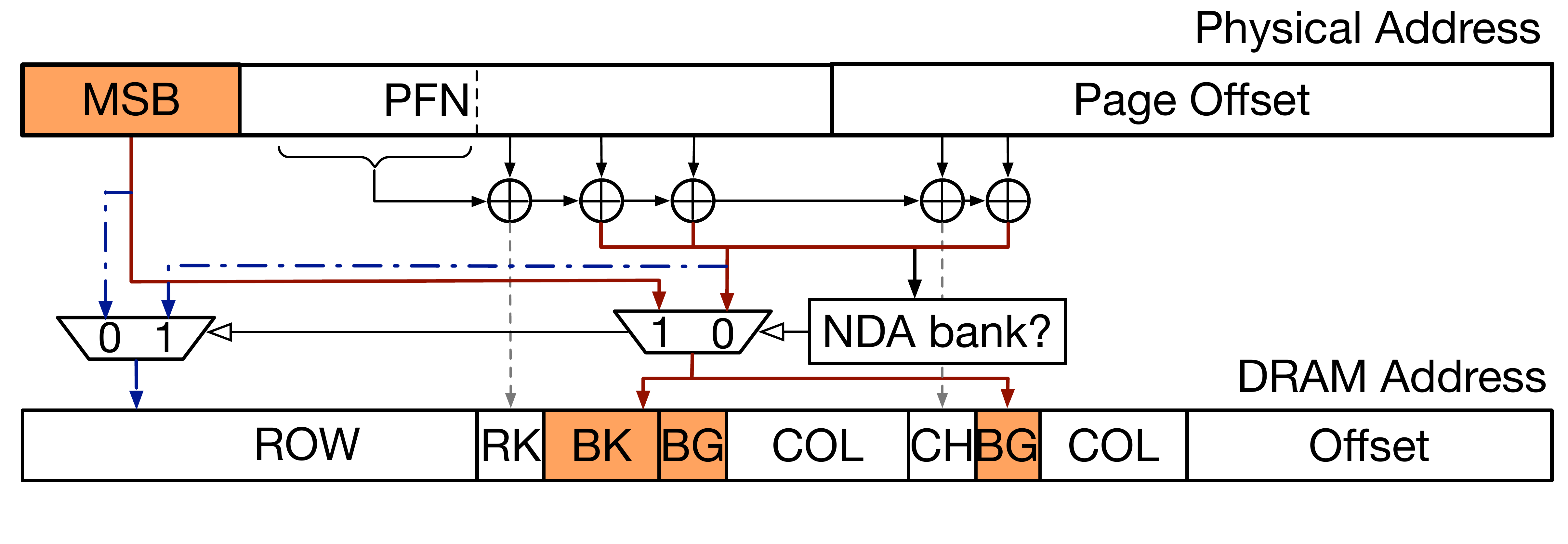}
				\label{fig:hashing_addr_map}
			}
		\end{minipage}
	\caption{Baseline and proposed host-side address mapping.}
	\label{fig:addr_map}
	\vspace*{-4mm}
\end{figure}

We rely on the NDA runtime and OS to use a combination of coarse-grain memory allocation and coloring to ensure all operands of an NDA instruction are interleaved across ranks the same way \meadd{and are thus local to a PE}. First, the runtime allocate memory for NDA operands such that they are aligned at the granularity of one DRAM row for each bank in the system which we call a \textit{system row} (e.g., 2MiB for a DDR4 1TiB system). For all the address interleaving mechanisms we are aware of~(\cite{pessl2016drama,liu2018get}), this ensures that NDA operands are locally aligned, as long as ranks are also kept aligned. To maintain rank alignment, we reply on OS page coloring to effect rank alignment. We explain this feature below using the Intel Skylake address mapping~\cite{pessl2016drama} as a concrete and representative interleaving mapping (\fig{fig:baseline_addr_map}).

In this mapping, rank and channel addresses are determined partly by the low-order bits that fall into the frame offset field and partly by the high-order bits that fall into the physical frame number (PFN) field. Frame offsets are kept the same because of the coarse-grain alignment. The OS colors allocations such that the PFN bits that determine rank and channel are aligned for a particular color; which physical address bits select ranks and channels can be reverse engineered if necessary~\cite{pessl2016drama}. The Chopim runtime indicates a \textit{shared color} when it requests memory from the OS and specifies the same color for all operands of an instruction. The runtime can use the same color for many operands to minimize copies needed for alignment. In our baseline system, there are 8 colors and each color corresponds to a shared region of memory of $4$GiB. Multiple regions can be allocated for the same process. Though we focus on one address mapping here, our approach works with any linear address mapping described in prior work~\cite{pessl2016drama,liu2018get} as well. 

Note that coarse-grain allocation is simple with the common buddy allocator if allocation granularity is also a system row, and can use optimizations that already exist for huge pages~\cite{yun2014palloc,kwon2016coordinated,gorman2004understanding}. The fragmentation overheads of coarse allocation are similar to those with huge pages and we find that they are negligible because coarse-grain NDA execution works best when processing long vectors.

\medskip
\noindent\textbf{\textit{Data layout across DRAM chips.}}
In the baseline system, each 4-byte word is striped across multiple chips, whereas in our approach each word is located in a single chip so that NDAs can access words from their local memory. Both the host and NDAs can access memory without copying or reformatting data (as required by prior work~\cite{farmahini2015nda}). Memory blocks still align with cache lines, so this layout change is not visible to software. \bcut{This layout precludes the critical word first optimization from DRAM, but recent work concludes the impact is minimal because the relative latency difference in current memory systems is very small (e.g.,~\cite{yoon2012dgms}).} Note that this data layout does not impact the host memory controller's ECC computation (e.g. Chip-kill~\cite{dell1997white}) because ECC protects only bits and not how they are interpreted. For NDA accesses, we rely on in-DRAM ECC with its limited coverage. We do not innovate in this respect and leave this problem for future work.

\subsection{Mitigating Frequent Read/Write Penalties}
\label{subsec:block_nda_write}

The basic memory access scheduling policy we use for Chopim is to always prioritize host memory requests, yet aggressively leverage unutilized rank bandwidth by issuing NDA requests whenever possible. That is, NDAs wait when incoming host requests are detected, but otherwise always issue their memory requests to maximize their bandwidth utilization and performance.
One potential problem is that an NDA request issued in one cycle may delay a host request that could have issued in one of the following cycles otherwise.

We find that NDAs infrequently issue row commands (ACT and PRE). We therefore prioritize host memory commands over any NDA row command to the same bank. This has negligible impact on NDA performance in our experiments.

We also find that read transactions of NDAs have only a small impact on following host commands. NDA write transactions, however, can have a large impact on host performance because of the read/write-turnaround penalties that they frequently require. While the host mitigates turnaround overhead by buffering operations with caches and write buffers~\cite{stuecheli2010virtual,ahn2006design}, the host and NDAs may interleave different types of transactions when accessing memory in parallel. We find that NDA writes interleaved with host reads degrade performance the most. \emph{As a solution,} we introduce two mechanisms to selectively throttle NDA writes. 


Our first mechanism throttles the rate of NDA writes by issuing them with a predefined probability. We call this mechanism \textit{stochastic NDA issue}. Before issuing a write transaction, the NDAs both detect if a rank is idle and flip a coin to determine whether to issue the write. By adjusting the coin weight, the performance of the host and NDAs can be traded off: higher write-issue probability leads to more frequent turnarounds while a lower probability throttles NDA progress. Deciding how much to throttle NDAs requires analysis or profiling, and we therefore propose a second approach as well.  

Our second approach does not require tuning, and we empirically find that it works well. In this \emph{next rank prediction} approach, the memory controller inhibits NDA write requests when more host read requests are expected; the controller stalls the NDA in lieu of providing an NDA write queue. In a packetized interface, the memory controller schedules both host and NDA requests and is thus aware of potential required turnarounds. The traditional memory interface, however, is more challenging as the host controller must explicitly signal the NDA controller to inhibit its write request. This signal must be sent ahead of the regular host transaction because of bus delays.

We use a very simple predictor that inhibits NDA write requests in a particular rank when the oldest outstanding host memory request to that channel is a read to that same rank. Specifically, the NDA controller examines the target rank of the oldest request in the host memory controller transaction queue. Then, it signals to the NDAs in that rank to stall their writes. For now, we assume that this information is communicated over a dedicated pin and plan to develop other signaling mechanisms that can piggyback on existing host DRAM commands at a later time. Our experiments with an FRFCFS~\cite{frfcfs} memory scheduler at the host show that this simple predictor works well and achieves performance that is comparable to a tuned stochastic issue approach.


\subsection{Partitioning into Host and Shared Banks}
\label{subsec:impl_bpart}

In addition to read/write-turnaround overheads, concurrent access also degrades performance by decreasing DRAM row access locality. When the host and NDAs interleave accesses to different rows of the same bank, frequent bank conflicts occur. To avoid this bank contention, we propose using bank partitioning to limit bank interference to only those memory regions that must concurrently share data between the NDAs and the host. This is particularly useful in colocation scenarios when only a small subset of host tasks utilize the NDAs. However, existing bank partitioning mechanisms~\cite{mi2010bankpark,jeong2012balancing,liu2012software} are incompatible with both huge pages and with sophisticated DRAM address interleaving schemes.

Bank partitioning relies on the OS to color pages where colors can be assigned to different cores or threads, or in our case, for banks isolated for the host and those that could be shared. The OS then maps pages of different color to frames that map to different banks.  
\fig{fig:baseline_addr_map} shows an example of a modern physical address to DRAM address mapping \cite{pessl2016drama}. One color bit in the baseline mapping belongs to the page offset field so prior bank partitioning schemes can, at best, be done at two-bank granularity. More importantly, when huge pages are used (e.g., 2MiB), this baseline mapping cannot be used to partition banks at all. 

To overcome this limitation, we propose a new interface that partitions banks into two groups---host-reserved and shared banks---with flexible DRAM address mapping and any page size. Specifically, our mechanism only requires that the most significant physical address bits are only used to determine DRAM row address, as is common in recent hash mapping functions, as shown in \fig{fig:hashing_addr_map} \cite{pessl2016drama}.

Without loss of generality, assume 2 banks out of 16 banks are reserved for the shared data. First, the OS splits the physical address space for host-only and shared memory region with the host-only region occupying the bottom of the address space: $0-\left(14\times\mathit{(bank\_capacity)}-1\right)$. The rest of the space (with the capacity of 2 banks) is reserved for the shared data and the OS does not use it for other purposes. This guarantees that the most significant bits (MSBs) of the address of host-only region are never b'111. In contrast, addresses in the shared space always have b'111 in their MSBs. 

The OS informs the memory controller that it reserved 2 banks (the top-most banks) for shared memory region. Host-only memory addresses are mapped to DRAM locations using any hardware mapping function, which is not exposed to software and the OS. The idea is then to remap addresses that initially fall into shared banks into the reserved address space that the host is not using. Additional simple logic checks whether the resulting DRAM address bank ID of the initial mapping is a reserved bank for shared region. If they are not, the DRAM address is used as is. If the DRAM address is initially mapped to one of the reserved banks, the MSBs and the bank bits are swapped. Because the MSBs of a host address are never b'1110 or b'1111, the final bank ID will be one of the host-only bank IDs. Also, because the bank ID of the initial mapping result is 14 or 15, the final address is in a row the host cannot access with the initial mapping and there is no aliasing. Note that the partitioning decision can be adjusted, but only if all affected memory is first cleared. 

\subsection{Tracking Global Memory Controller State}
\label{subsec:track_gstate}

\begin{figure}[t!]
\centering
        \includegraphics[width=0.42\textwidth]{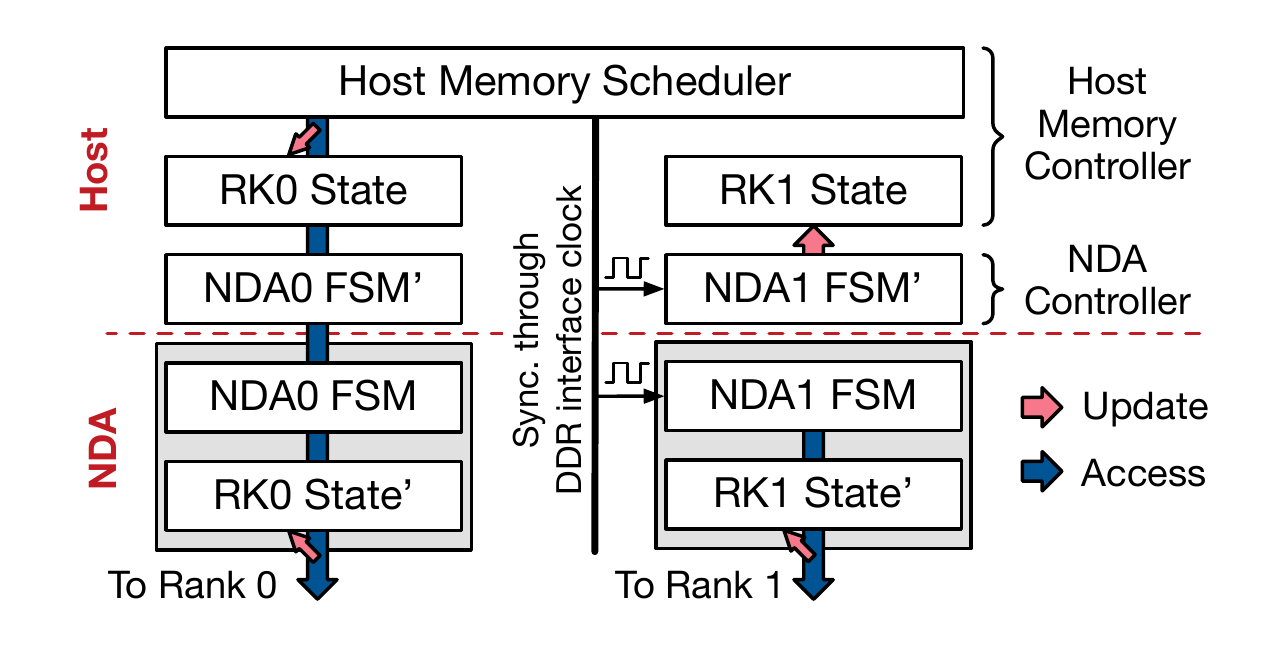}
	\caption{Global MC state tracking when the host (left) and NDAs (right) issues memory commands. The replicated FSMs are synchronized by using the DDR interface clock.}
	\label{fig:repl_fsm}
	\vspace*{-2mm}
\end{figure}

Unlike conventional systems, Chopim also enables an architecture that has two memory controllers (MCs) managing the bank and timing state of each rank. This is the case when the host continues to directly manage memory even when the memory itself is enhanced with NDAs. This requires coordinating rank state information between controllers. \fig{fig:repl_fsm} shows how MCs on both sides of a memory channel track global memory controller state. Information about host transactions is easily obtained by the NDA MCs as they can monitor incoming transactions and update the state tables accordingly (left). However, the host MC cannot track all NDA transactions due to command bandwidth limits.

To solve this problem, we replicate the finite-state machines (FSMs) of NDAs and place them in the host-side NDA controller. When an NDA instruction is launched, the FSMs on both sides are synchronized. We rely on the already-synchronized DDR interface clock for FSM synchronization. Whenever an NDA memory transaction is issued, the host-side FSM also updates the state table in the host MC without communicating with the NDAs (right). If a host transaction blocks NDA transactions in one of the ranks, that transaction will be visible to both FSMs. Replicated FSMs track the NDA write buffer occupancy and detect when the write-buffer draining starts and ends to trigger write throttling. The area and power overhead of replicating FSMs are negligible (40-byte microcode store and 20-byte state registers per rank (i.e., per NDA)).
\meadd{\emph{Our evaluation uses this approach to enable a DDR4-based NDA-enabled main memory and all our experiments rely on this.}}

%% file: tex/collaboration.tex

\section{Host-NDA Collaboration}
\label{sec:collaboration}

In this section, we describe a case study to show the potential of concurrent host-NDA execution by collaboratively processing the same data. Our case study shows how to partition \meadd{an ML training task between the host and NDAs such that each processor leverages its strengths. As is common to training and many data-processing tasks, the vast majority of shared data is read-only, simplifying parallelism.}
\medel{Also, our case study is a good example since infrequent and low-overhead operations are required to maintain coherence while the host and NDAs can independently access large and shared read-only data of which access time dominates the overall execution time.}

\begin{figure}[t!]
\centering
	\includegraphics[width=0.48\textwidth]{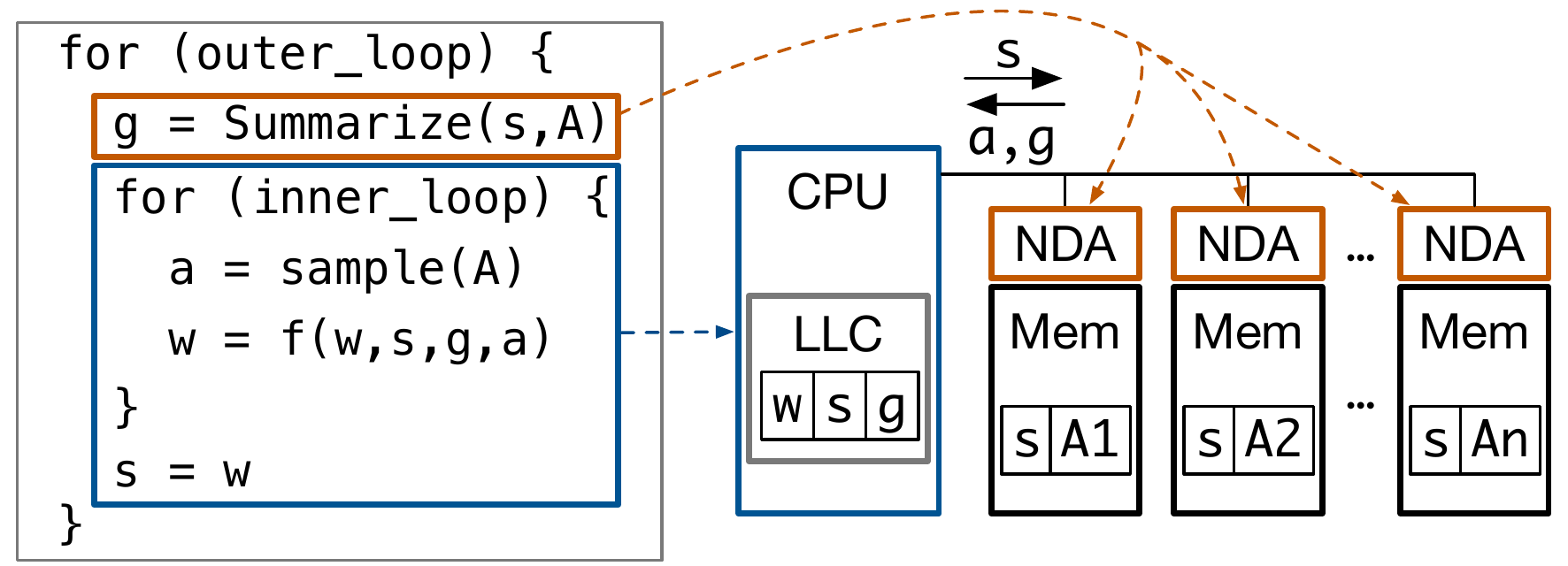}
	\caption{Collaboration between host and NDAs in SVRG.}
	\label{fig:svrg}
	\vspace*{-4mm}
\end{figure}

We use the machine-learning technique of logistic regression with stochastic variance reduced gradient (SVRG)~\cite{johnson2013accelerating} as our case study. \medel{SVRG is a machine learning technique that enables faster convergence by reducing variance introduced by sampling.} \fig{fig:svrg} shows a simplified version of SVRG and the opportunity for collaboration.
\meadd{The algorithm consists of two main tasks within each outer-loop iteration. First, the entire large input matrix \textit{A} is \emph{summarized} into a single vector \textit{g} (see \fig{fig:impl_avg_grad_example_code} for pseudocode). This vector is used as a correction term when updating the model in the second task. This second task consists of multiple inner-loop iterations. In each inner-loop iteration the learned model \textit{w} based on a randomly-sampled vector \textit{a} from the large input matrix \textit{A}, the correction term \textit{g}, and a stored model \textit{s}, which is updated at the end of the outer-loop iteration. }
\medel{
A large input matrix, \textit{A}, is evenly partitioned into multiple tiles and stored in memory. In every inner-loop iteration, the host samples a random element \textit{a} within \textit{A} to update the learned model \textit{w}. Other than the large input, other data (\textit{w, s,} and \textit{g}) takes advantage of the CPU caches. The tight inner loop is therefore ideally suited for high-end CPU execution.
}

The first task is an excellent match for the NDAs. 
\medel{The SVRG algorithm periodically calculates a correction term, \textit{g}, by \textit{summarizing} the entire input data (example code in \fig{fig:impl_avg_grad_example_code}). Because} The summarization operation is simple, exhibits little reuse, and traverses the entire large input data. \medel{, it is ideally suited for the NDAs The term \textit{g} is used for correcting error in the host workload, \textit{f}. With Chopim,}
In contrast, the second task with its tight inner loop is well suited for the host. The host can maximally exploit locality captured by its caches while NDAs can leverage their high  bandwidth for accessing the entire input data \textit{A}. Note that in SVRG, an \textit{epoch} refers to the number of inner loop iterations. 

The main tradeoff in SVRG is as follows. When summarization is done more frequently, the quality of the correction term increases and, consequently, the per-step convergence rate increases. On the other hand, the overhead of summarization also increases when it is performed more frequently, which  offsets the improved convergence rate. Therefore, the \textit{epoch} hyper-parameter, which determines the frequency of summarization, should be carefully selected to optimize this tradeoff.

\medskip
\noindent\textbf{\textit{Delayed-Update SVRG.}}
As Chopim enables concurrent access from the host and NDAs, we explore an algorithm change to leverage collaborative parallel processing. Instead of alternating between the summarization and model update tasks, we run them in parallel on the host and NDAs. Whenever the NDAs finish computing the correction term, the host and NDAs exchange the correction term and the most up-to-date weights before continuing parallel execution. While parallel execution is faster, it results in using stale \textit{s} and \textit{g} values from one epoch behind. The main tradeoff in \textit{delayed-update SVRG} is that per-iteration time is improved by overlapping execution, whereas convergence rate per iteration degrades due to the staleness. Similar tradeoffs have been observed in prior work \cite{bengio2003neural,langford2009slow,recht2011hogwild,dean2012large}. \meadd{We later show that delayed-update SVRG can converge in $40\%$ less time than when serializing the two main SVRG tasks.}

To avoid races for \textit{s} and \textit{g} in this delayed-update SVRG, we maintain private copies of these small variables and use a memory fence that guarantees completion of DRAM writes after the data-exchange step (which the runtime coordinates with polling). Note that we bypass caches when accessing data produced/consumed by NDAs during the data-exchange step. Since \textit{s} and \textit{g} are small and copied infrequently, the overheads are small and amortized over numerous NDA computations. Whether delayed updates are used or not, the host and NDAs share the large data, \textit{A}, without copies.


%% file: tex/implementation.tex

\section{Runtime and API}
\label{sec:implementation}

\begin{figure}[t!]
\centering
	\includegraphics[width=0.46\textwidth]{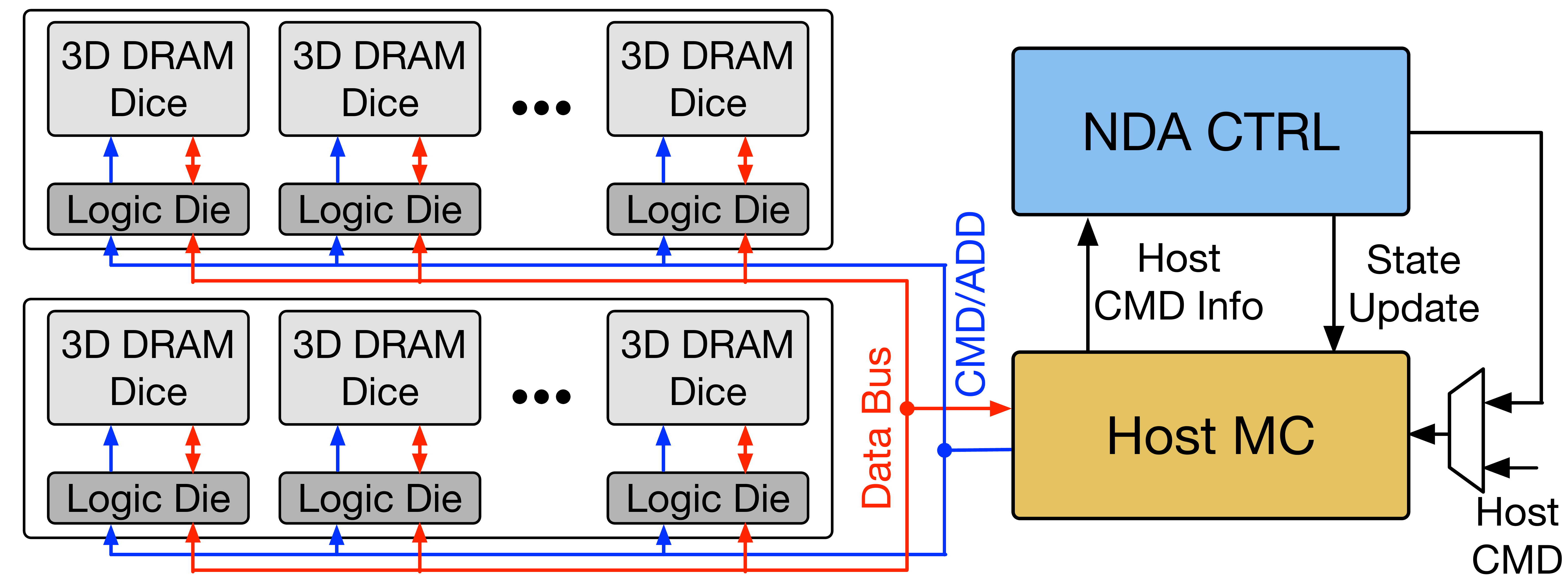}
	\caption{Overview of NDA architecture.}
	\label{fig:impl_overview}
	\vspace*{-4mm}
\end{figure}

\begin{figure}[t!]
\centering
	\includegraphics[width=0.38\textwidth]{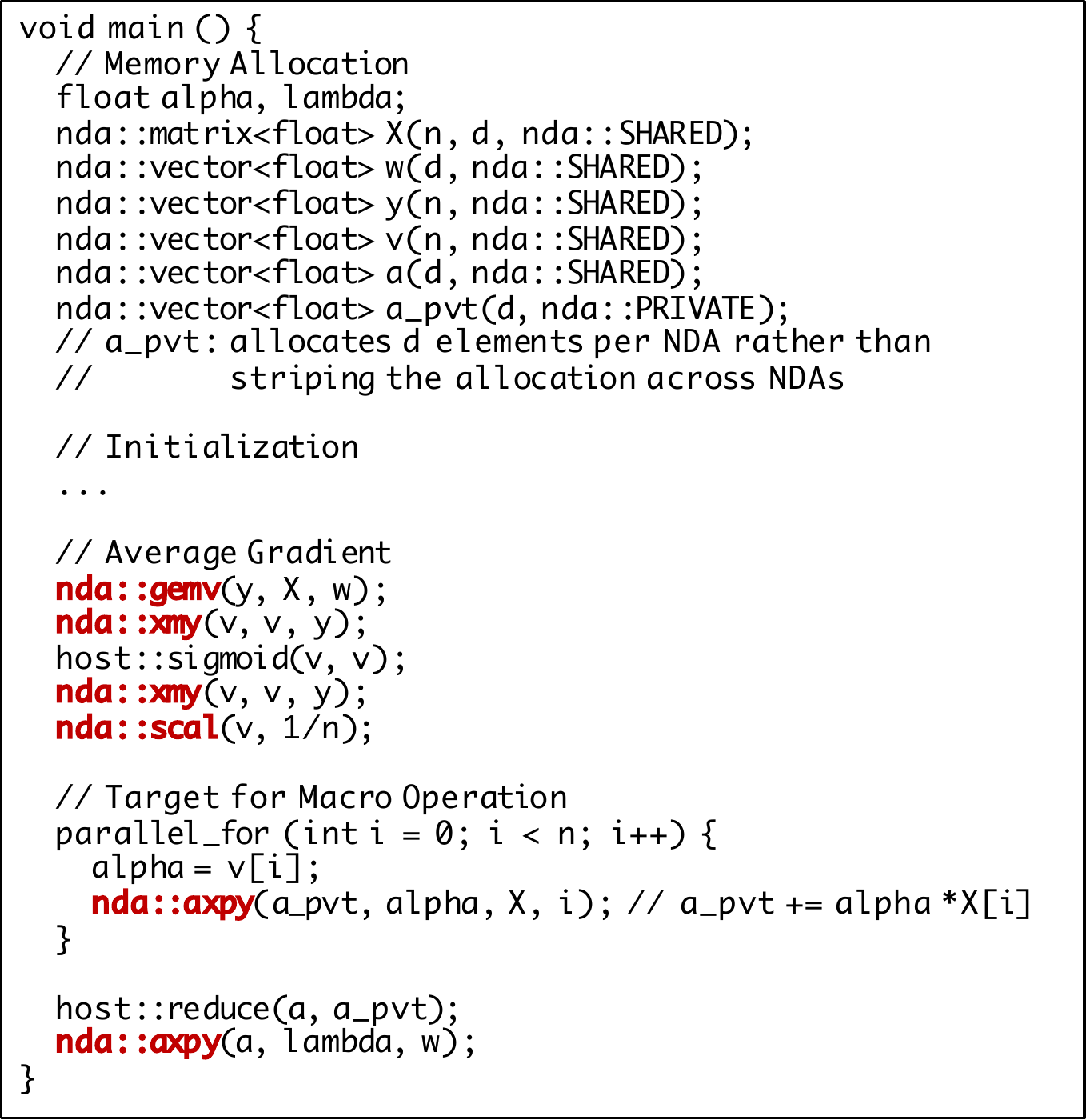}
	\caption{Average gradient example code. This code corresponds to \textit{summarization} in SVRG (see Section \ref{sec:collaboration}).}
	\label{fig:impl_avg_grad_example_code}
\end{figure}

Chopim is general and helps whenever host/NDA concurrent access is needed. To make the explanations and evaluation concrete, we use an exemplary \meadd{interface} design as discussed below and summarized in \fig{fig:impl_overview}. Command and address signals pass through the NDA memory controllers so that they can track host rank state. Processing elements (PEs) in the logic die access data by using their local NDA memory controller (\fig{fig:baseline_nda}).
\bcut{We propose a similar API as other C++ math libraries~\cite{sanderson2010armadillo,jacob2013eigen,iglberger2012high} for the example use case of accelerating linear algebra operations.} \fig{fig:impl_avg_grad_example_code} shows example usage of our API for computing the average gradient used in the summarization task of SVRG. 

The Chopim runtime system manages memory allocations and launches NDA operations. NDA operations are blocking by default, but can also execute asynchronously. If the programmer calls an NDA operation with operands from different shared regions (colors), the runtime system inserts appropriate data copies. We envision a just-in-time compiler that can identify such cases and more intelligently allocate memory and regions to minimize copies. For this paper, we do not implement such a compiler. Instead, programs are written to directly interact with a runtime system that is implemented within the simulator.

NDAs operate directly on DRAM addresses and do not perform address translation. To launch an operation, the runtime (with help from the OS) translates the origin of each operand into a physical address, which is then communicated  \meadd{along with a bound} to the NDAs by the NDA controller. The runtime is responsible for splitting a single API call into multiple primitive NDA operations. The NDA operations themselves proceed through each operand with a regular access pattern implemented as microcode in the hardware\meadd{, which also checks the bound for protection}. \bcut{DRAM addresses are computed by following the same physical-to-DRAM mapping function used by the host memory controller (\sect{subsec:impl_data_layout}).}

\medskip
\noindent\textbf{\textit{Optimization for Load-Imbalance.}}
Load imbalance occurs when the host does not access ranks uniformly over short periods of time. The AXPY operation (launched repeatedly within the loop shown in \fig{fig:impl_avg_grad_example_code}) is short and non-uniform access by the host leads to load imbalance among NDAs. A blocking operation waits for \emph{all} NDAs to complete before launching the next AXPY, which reduces performance. 
Our API provides asynchronous launches similar to CUDA streams \ykadd{or OpenMP parallel \texttt{for} with a \texttt{nowait} clause \cite{dagum1998openmp}}. Asynchronous launches can overlap AXPY operations from multiple loop iterations. Any load imbalance is then only apparent when the loop ends. Over such a long time period, load imbalance is much less likely. We implement asynchronous launches using \emph{macro NDA operation}. An example of a macro operation is shown in the loop of \fig{fig:impl_avg_grad_example_code} and is indicated by the \texttt{parallel\_for} annotation. 

\bcut{
\medskip
\noindent\textbf{\textit{Exploiting Inter-Iteration Locality.}}
Each NDA PE includes a small 1 KB scratchpad memory (sized equal to a row buffer within the DRAM chip). The runtime leverages this to reorder operations within macro NDA operations. In the AXPY macro operation example, inter-iteration locality exists for vector ${\vec{a}}$. If ${\vec{a}}$ does not fit in the scratchpad memory, matrix ${X}$ is decomposed in the column direction and operations are launched for one column group after another. The locality captured by the scratchpad eliminates writing intermediate results back into DRAM. This also reduces write interference (write-to-read and read-to-write turnaround times).
}

\medskip
\noindent\textbf{\textit{Launching NDA Operations.}}
\label{subsec:impl_launch}
NDA operations are launched similarly to Farmahini et al.~\cite{farmahini2015nda}. A  memory region is reserved for accessing control registers of NDAs. NDA packets access the control registers and launch operations. Each packet is composed of the type of operation, the base addresses of operands, the size of data blocks, and scalar values required for scalar-vector operations. On the host side, the \textit{NDA controller} plays two main roles. First, it accepts acceleration requests, issues commands to the NDAs in the different ranks (in a round-robin manner), and notifies software when a request completes. Second, it extends the host memory controller to coordinate actions between the NDAs and host memory controllers and enables concurrent access. It maintains the replicated FSMs using its knowledge of issued NDA operations and the status of the host memory controller. \bcut{The NDA controller is also responsible for throttling specific NDAs if necessary to maintain host performance.}

\medskip
\noindent\textbf{\textit{Execution Flow of a Processing Element.}}
\label{subsec:impl_pe}

\begin{figure}[t!]
\centering
	\includegraphics[width=0.42\textwidth]{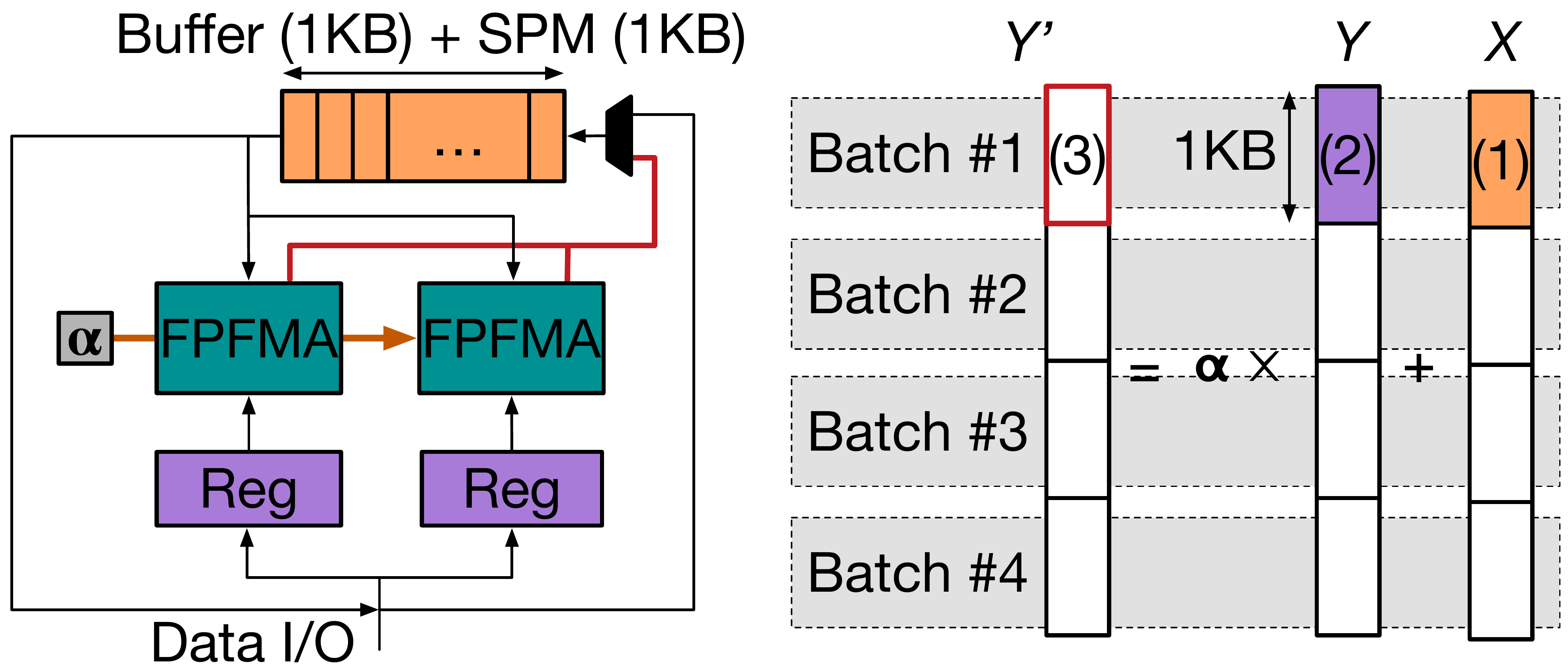}
	\caption{PE architecture and execution flow of AXPY.}
	\label{fig:eflow_axpy}
	\vspace*{-4mm}
\end{figure}

Our exemplary PE is composed of two floating-point fused multiply-add (FPFMA) units, 5 scalar registers (up to 3 operand inputs and 2 for temporary values), a 1KB buffer for accessing memory, and the 1KB scratchpad memory. The memory access granularity is 8B per chip and the performance of the two FPFMAs per chip matches this data access rate. PEs may be further optimized to support lower-precision operations or specialized for specific use cases, but we do not explore these in this paper as we focus on the new capabilities of Chopim rather than NDA in general.

\fig{fig:eflow_axpy} shows the execution flow of a PE when executing the AXPY operation. Each vector is partitioned into 1KB batches, which is the same size as DRAM page size per chip. To maximize bandwidth utilization, the vector ${X}$ is streamed into the buffer. Then, the PE opens another row, reads two elements (8 bytes) of vector ${Y}$, and stores them to FP registers. While the next two elements of ${Y}$ are read, a fused multiply-add (FMA) operation is executed. The result is stored back into the buffer and execution continues such that the read-execute-write operations are pipelined. After the result buffer is filled, the PE either writes results back to memory or to the scratchpad. This flow for one 1KB batch is repeated over the rest of the batches. This entire process is stored in PE microcode as the AXPY operation. \meadd{Other operations (coarse or fine grained) are similarly stored and processed from microcode.}

\bcut{
Note that, if we only have one NDA bank, changing the access order of two input vectors degrades the performance of AXPY. This is because if vector ${Y}$ is read first and vector ${X}$ next, the row for vector ${Y}$ is closed before it is updated, whereas the reverse order will guarantee the row remains opened. Also, one optimization is to close a row right after accessing the last column of the row when the row is no longer being used. In AXPY, we always apply this optimization for ${X}$ whereas only apply for ${Y}$ after writing is done.

Other NDA operations (\tab{tab:nda_ops}) follow a similar execution flow. NRM2 is a dot-product of one vector and itself. Therefore, the input to the PE should be written to the buffer and to registers at the same time. NRM2 and DOT require reductions at the end since two FPFMAs operate separately on their own accumulators; the reductions are performed by the runtime system on the host. The input of the SCAL operation is stored directly into the register and the results are written to the buffer.
} 

\medskip
\noindent\textbf{\textit{Inter-PE Communication.}}
\meadd{NDAs are only effective when they use memory-side bandwidth to amplify that of the host. In the DIMM- and chip-based NDAs, which we target in this paper, general inter-PE communication is therefore equivalent to communicating with the host. Communication in applications that match this NDA architecture are primarily needed for replicating data to localize operands or for global reduction operations, which follow local per-PE reductions.} 
\medel{There are two types of communication in our case study: data replication and reduction.} In both communication cases, a global view of data layout is needed and, therefore, we enable communication only through the host. For instance, after the macro operation in \fig{fig:impl_avg_grad_example_code}, \meadd{a global reduction of the PE private copies (\texttt{a\_pvt})  accumulates the data for the final result (\texttt{a}). The reduced result is used by the following NDA operation, requiring replication communication for its data layout has to meet NDA locality requirements with the other NDA operand (\texttt{w}). Though communicating through the host is expensive, our coarse-grained NDA operations amortize infrequent communication overhead. Importantly}, since this communication can be done as normal DRAM accesses by the host, no change on the memory interface is required.


%% file: tex/methodology.tex

\section{Methodology}
\label{sec:method}

Table \ref{tab:eval_config} summarizes our system configuration, DRAM timing parameters, energy components, benchmarks, and machine learning configurations. For bank partitioning, we reserve one bank per rank for NDAs and the rest for the host. We use Ramulator \cite{kim2016ramulator} as our baseline DRAM simulator and add the NDA memory controllers and PEs to execute the NDA operations. We modify the memory controller to support the Skylake address mapping~\cite{pessl2016drama} and our bank partitioning and data layout schemes. To simulate concurrent host accesses, we use gem5 \cite{binkert2011gem5} with Ramulator. We choose host applications that have various memory intensity from the \textit{SPEC2006} \cite{henning2006spec} and \textit{SPEC2017} \cite{panda2018wait} benchmark suites and form 9 different application mixes with different combinations (Table \ref{tab:eval_config}). Mix0 and mix8 represent two extreme cases with the highest and lowest memory intensity, respectively. Only mix0 is run with 8 cores to simulate under-provisioned bandwidth while other mixes use 4 cores to simulate a more realistic scenario. \bcut{Since Chopim is important only when the host processor and PEs concurrently try to access memory, we only show the results of benchmarks with medium and high memory intensity. We also ran simulations with low memory intensity benchmarks and the performance impact due to contention is negligible.} For the NDA workloads, we use DOT and COPY operations to show the impact of extremely low and high write intensity. We use the average gradient kernel (\fig{fig:impl_avg_grad_example_code}) to evaluate collaborative execution. The performance impact of other NDA applications falls between DOT and COPY and is well represented by SVRG {\cite{johnson2013accelerating}}, conjugate gradient (CG) {\cite{jacob2013eigen}} and streamcluster (SC) {\cite{pisharath2005nu}}.

For the host workloads, we use Simpoint \cite{hamerly2005simpoint} to find representative program phases and run each simulation until the instruction count of the slowest process reaches 200M instructions. If an NDA workload completes while the simulation is still running, it is relaunched so that concurrent access occurs throughout the simulation time. Since the number of instructions simulated is different, we measure instructions per cycle (IPC) for the host performance. To show how well the NDAs utilize bandwidth, we show bandwidth utilization and compare with the idealized case where NDAs can utilize all the idle rank bandwidth. 


We estimate power with the parameters in Table~\ref{tab:eval_config}. We use CACTI 6.5~\cite{muralimanohar2009cacti} for the dynamic and leakage power of the PE buffer. A sensitivity study for PE parameters exhibits that their impact is negligible. We use CACTI-3DD~\cite{chen2012cacti} to estimate the power and energy of 3D-stacked DRAM and CACTI-IO~\cite{jouppi2015cacti} to estimate   DIMM power and energy.

\begin{table}[!t]
	\centering
	\noindent\resizebox{\linewidth}{!}{
		\tabulinesep=0.6mm
		\begin{tabu}{c|c|[1.0pt]c|c|c}
			\hline
			\rowfont{\normalsize}
            \multicolumn{5}{c}{System configuration} \tabularnewline
			\hline
      Processor & \multicolumn{4}{c}{\makecell{4-core OoO x86 (8 cores for mix0), 4GHz, Fetch/Issue width (8), \\ LSQ (64), ROB (224)}} \tabularnewline
			\hline 
      NDA & \multicolumn{4}{c}{\makecell{one PE per chip, 1.2GHz, fully pipelined, write buffer (128) (Section {\ref{sec:implementation}})}} \tabularnewline
			\hline 
			TLB & \multicolumn{4}{c}{I-TLB:64, D-TLB:64, Associativity (4)} \tabularnewline
			\hline 
			L1 & \multicolumn{4}{c}{\makecell{32KB, Associativity (L1I: 8, L1D: 8), LRU, 12 MSHRs}} \tabularnewline
			\hline 
			L2 & \multicolumn{4}{c}{\makecell{256KB, Associativity (4), LRU, 12 MSHRs}} \tabularnewline
			\hline
			LLC & \multicolumn{4}{c}{\makecell{8MB, Associativity (16), LRU, 48 MSHRs, Stride prefetcher}} \tabularnewline
			\hline 
			DRAM & \multicolumn{4}{c}{\makecell{DDR4, 1.2GHz, 8Gb, x8, 2channels $\times$ 2ranks, \\
			FR-FCFS, 32-entry RD/WR queue, Open policy, \\
			Intel Skylake address mapping \cite{pessl2016drama}}} \tabularnewline
			\hline
			\arrayrulecolor{white}\hline
			\arrayrulecolor{white}\hline
			\arrayrulecolor{white}\hline
			\arrayrulecolor{black}\hline
			\rowfont{\normalsize}
            \multicolumn{5}{c}{DRAM timing parameters} \tabularnewline
			\hline

			\multicolumn{5}{c}{\makecell{tBL=4, tCCDS=4, tCCDL=6, tRTRS=2, tCL=16, tRCD=16,\\
			tRP=16, tCWL=12, tRAS=39, tRC=55, tRTP=9, tWTRS=3,\\
      tWTRL=9, tWR=18, tRRDS=4, tRRDL=6, tFAW=26}} \tabularnewline
			\hline 

			\hline
			\arrayrulecolor{white}\hline
			\arrayrulecolor{white}\hline
			\arrayrulecolor{white}\hline
			\arrayrulecolor{black}\hline
			\rowfont{\normalsize}
            \multicolumn{5}{c}{Energy Components} \tabularnewline
			\hline

			\multicolumn{5}{c}{\makecell{Activate energy: 1.0nJ, PE read/write energy: 11.3pJ/b, \\
			host read/write energy: 25.7pJ/b, PE FMA: 20pJ/operation, \\
			PE buffer dynamic: 20pJ/access, PE buffer leakage power: 11mW \\
			(Energy/power of scratchpad memory is same as PE buffer)}} \tabularnewline
			\hline 
			
			\hline
			\arrayrulecolor{white}\hline
			\arrayrulecolor{white}\hline
			\arrayrulecolor{white}\hline
			\arrayrulecolor{black}\hline
			\rowfont{\normalsize}
            \multicolumn{4}{c|}{Benchmarks} & MPKI\tabularnewline
			\hline
      mix0 & \multicolumn{3}{c|}{\makecell{mcf\_r:lbm\_r:omnetpp\_r:gemsFDTD\\
      bwaves:milc:soplex:leslie3d}} & \makecell{H:H:H:H\\
      H:M:M:M}\tabularnewline
			\hline 
			mix1 & \multicolumn{3}{c|}{mcf\_r:lbm\_r:omnetpp\_r:gemsFDTD} & H:H:H:H\tabularnewline
			\hline 
			mix2 & \multicolumn{3}{c|}{mcf\_r:lbm\_r:gemsFDTD:soplex} & H:H:H:H\tabularnewline
			\hline 
			mix3 & \multicolumn{3}{c|}{lbm\_r:omnetpp\_r:gemsFDTD:soplex} & H:H:H:H\tabularnewline
			\hline 
			mix4 & \multicolumn{3}{c|}{omnetpp\_r:gemsFDTD:soplex:milc} & H:H:H:M\tabularnewline
			\hline 
			mix5 & \multicolumn{3}{c|}{gemsFDTD:soplex:milc:bwaves\_r} & H:H:M:M\tabularnewline
			\hline 
			mix6 & \multicolumn{3}{c|}{soplex:milc:bwaves\_r:leslie3d} & H:M:M:M\tabularnewline
			\hline 
			mix7 & \multicolumn{3}{c|}{milc:bwaves\_r:astar:cactusBSSN\_r} & M:M:M:M\tabularnewline
			\hline 
      mix8 & \multicolumn{3}{c|}{leslie3d:leela\_r:deepsjeng\_r:xchange2\_r} & M:L:L:L\tabularnewline
			\hline

      \hline
			\arrayrulecolor{white}\hline
			\arrayrulecolor{white}\hline
			\arrayrulecolor{white}\hline
			\arrayrulecolor{black}\hline
			\rowfont{\normalsize}
            \multicolumn{5}{c}{NDA Kernels} \tabularnewline
			\hline

      \multicolumn{5}{c}{\makecell{NDA basic operations (Table {\ref{tab:nda_ops}}), SVRG (details below), \\
      CG (16K ${\times}$ 16K), and SC (2M ${\times}$ 128)}} \tabularnewline
			\hline

			\hline
			\arrayrulecolor{white}\hline
			\arrayrulecolor{white}\hline
			\arrayrulecolor{white}\hline
			\arrayrulecolor{black}\hline
			\rowfont{\normalsize}
            \multicolumn{5}{c}{Machine Learning Configurations} \tabularnewline
			\hline

			\multicolumn{5}{c}{\makecell{Logistic regression with ${\ell2}$-regularization (10-class classification), ${\lambda}$=1e-3,\\
			learning rate=best-tuned, momentum=0.9, dataset=cifar10 (50000 ${\times}$ 3072)}} \tabularnewline
			\hline 
		\end{tabu}
	}
	\caption{Evaluation parameters.}
	\label{tab:eval_config} 
	\vspace*{-5mm}
\end{table}



%% file: tex/evaluation.tex

\section{Evaluation}
\label{sec:evaluation}


We present evaluation results for the various Chopim mechanisms, analyzing:
(1) the benefit of coarse-grain NDA operations; (2) how bank partitioning improves NDA performance; (3) how stochastic issue and next-rank prediction mitigate read/write turnarounds; (4) the impact of NDA workload write intensity and load imbalance; (5) how Chopim compares with rank partitioning; (6) the benefits of collaborative and parallel CPU/NDA processing; and (7) energy efficiency.
\meadd{All results rely on the replicated FSM to enable using DDR4.}

\begin{figure}[t!]
\centering
	\includegraphics[width=0.48\textwidth]{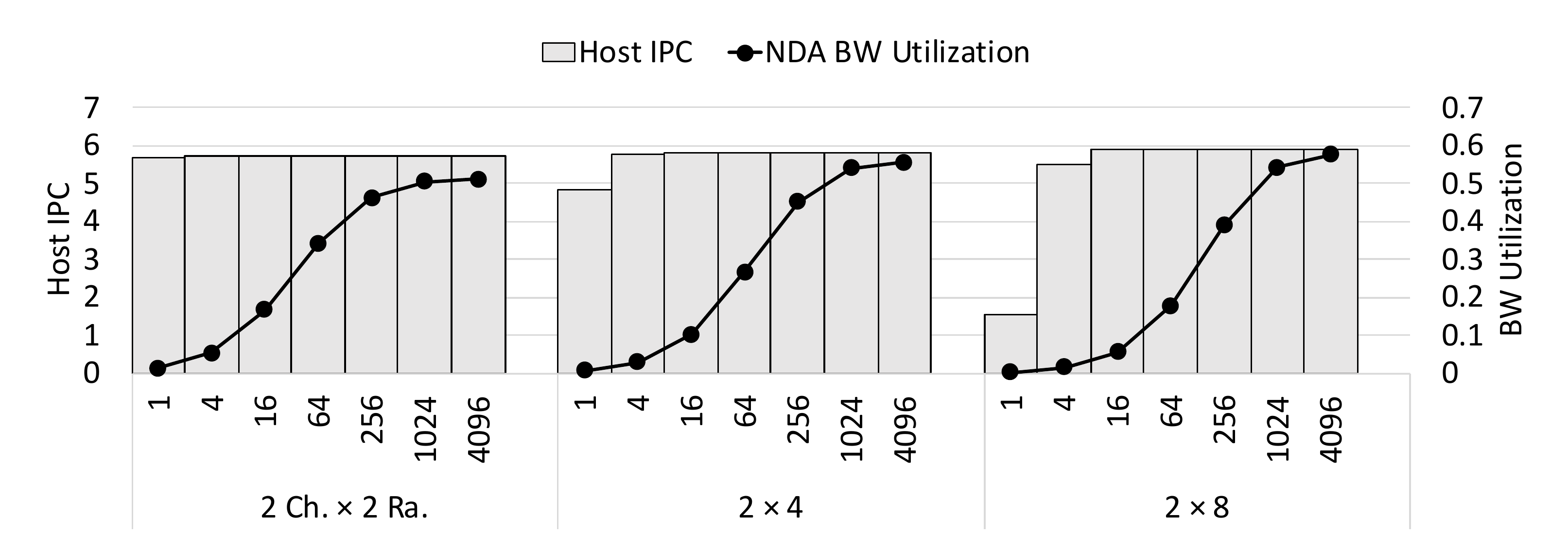}
	\caption{Impact of coarse-grain NDA operations. (X-axis: the number of cache blocks accessed per NDA instruction.)}
	\label{fig:cgnda}
	\vspace*{-4mm}
\end{figure}

\medskip
\noindent\textbf{\textit{Coarse-grain NDA Operation.}}
\fig{fig:cgnda} demonstrates how overhead for launching NDA instructions can degrade performance of the host and NDAs as rank count increases. To prevent other factors, such as bank conflicts, bank-level parallelism, and load imbalance from affecting performance, we use our BP mechanism, the NRM2 operation (because we can precisely control its granularity), and asynchronous launch. We run the most memory-intensive application mix (mix1) on the host. When more CBs are processed by each NDA instruction, contention between host transactions and NDA instruction launches decreases and performance of both improves. In addition, as the number of ranks grows, contention becomes severe because more NDA instructions are necessary to keep all NDAs busy. These results show that our data layout that enables coarse-grain NDA operation is beneficial, especially in concurrent access situation.

\medskip
\noindent\fbox{\begin{minipage}{0.46\textwidth}
    Takeaway 1: 
    Coarse-grain NDA operations are crucial for mitigating contention on the host memory channel.
\end{minipage}}

\begin{figure}[t!]
\centering
	\includegraphics[width=0.48\textwidth]{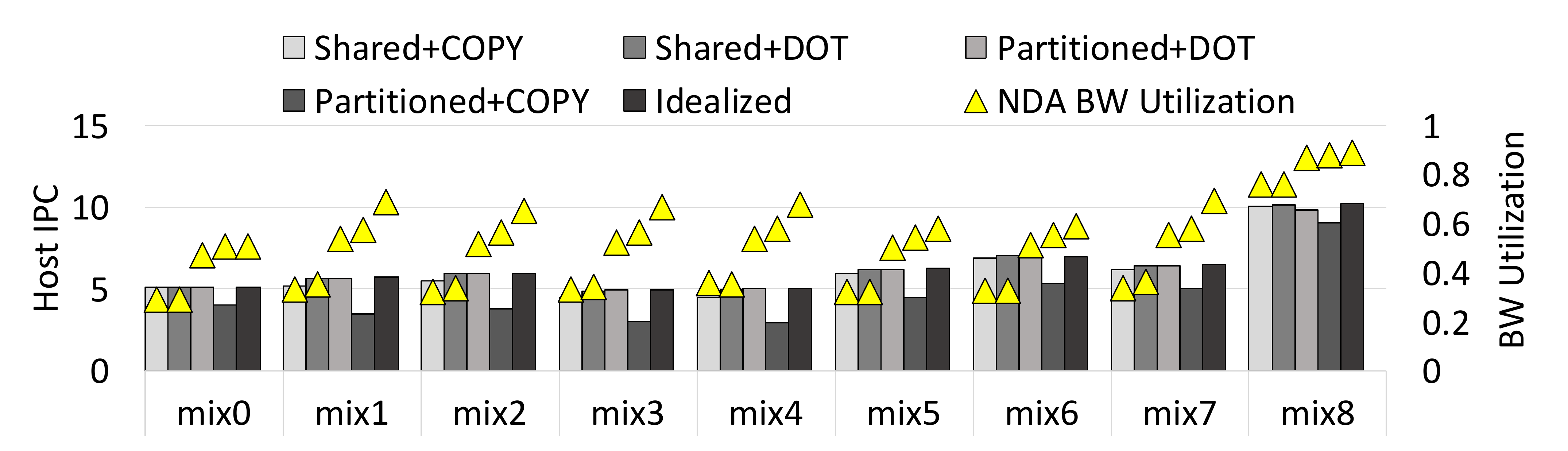}
	\caption{Concurrent access to different memory regions.}
	\label{fig:eval_bpart_vs_bshar}
\end{figure}

\medskip
\noindent\textbf{\textit{Impact of Bank Partitioning.}}
\fig{fig:eval_bpart_vs_bshar} shows performance when banks are shared or partitioned between the host and NDAs\mereplace{}{ which access different data}. We emphasize the impact of write intensity of NDA operations by running the extreme DOT (read intensive) and COPY (write intensive) operations. \meadd{While not shown, SVRG falls roughly in the middle of this range.} We compare each memory access mode with an idealized case where we assume the host accesses memory without any contention and NDAs can leverage all the idle rank bandwidth without considering transaction types and other overheads. 

Overall, accelerating the read-intensive DOT with concurrent host access does not affect host performance significantly even with our aggressive approach. However, contention with the shared access mode significantly degrades NDA performance. This is because of the extra bank conflicts caused by interleaving host and NDA transactions. On the other hand, accelerating the write-intensive COPY degrades host performance. This happens because, in the write phase of NDAs when the NDA write buffer drains, the host reads are blocked while NDAs keep issuing write transactions due to long write-to-read turnaround time. To mitigate this problem, we show the impact of our write throttling mechanisms below. Note that host performance of mix0 is the lowest, despite its doubled core count, because contention for LLC increases and memory performance dominates overall performance.

\medskip
\noindent\fbox{\begin{minipage}{0.46\textwidth}
    Takeaway 2: Bank partitioning increases row-buffer locality and substantially improves NDA performance, especially for read-intensive NDA operations.
\end{minipage}}

\begin{figure}[t!]
\centering
	\includegraphics[width=0.48\textwidth]{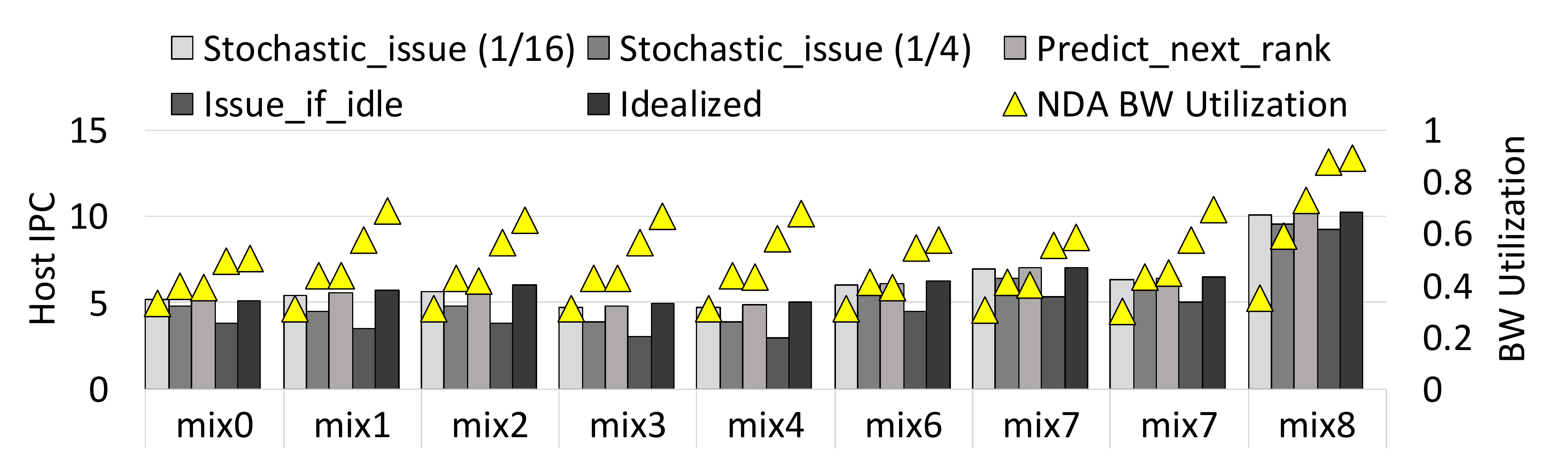}
	\caption{Stochastic issue and next-rank prediction impact.}
	\label{fig:eval_mech_nda_write}
	\vspace*{-4mm}
\end{figure}

\medskip
\noindent\textbf{\textit{Mitigating NDA Write Interference.}}
\fig{fig:eval_mech_nda_write} shows the impact of mechanisms for write-intensive NDA operations. In this experiment, the most write-intensive operation, COPY, is executed by NDAs and the mechanisms are applied only during the write phase of NDA execution. Stochastic issue is used with two probabilities, 1/4 and 1/16, which clearly shows the host-NDA performance tradeoff compared to next-rank prediction. 

For stochastic issue, the tradeoff between host and NDA performance is clear. If NDAs issue with high probability, host performance degrades. The appropriate issue probability can be chosen with heuristics based on host memory intensity though we do not explore this in this paper. On the other hand, the next-rank prediction mechanism shows slightly better behavior than the stochastic approach. Compared to stochastic issue with probability 1/16, both host and NDA performance are higher. Stochastic issue extends the tradeoff range and does not require signaling. \meadd{We use the robust next-rank prediction approach for the rest of the paper.}

\medskip
\noindent\fbox{\begin{minipage}{0.46\textwidth}
Takeaway 3: Throttling NDA writes mitigates the large impact of read/write turnaround interference on host performance; next-rank prediction is robust and effective while stochastic issue does not require additional signaling. 
\end{minipage}}

\begin{figure}[t!]
\centering
	\includegraphics[width=0.46\textwidth]{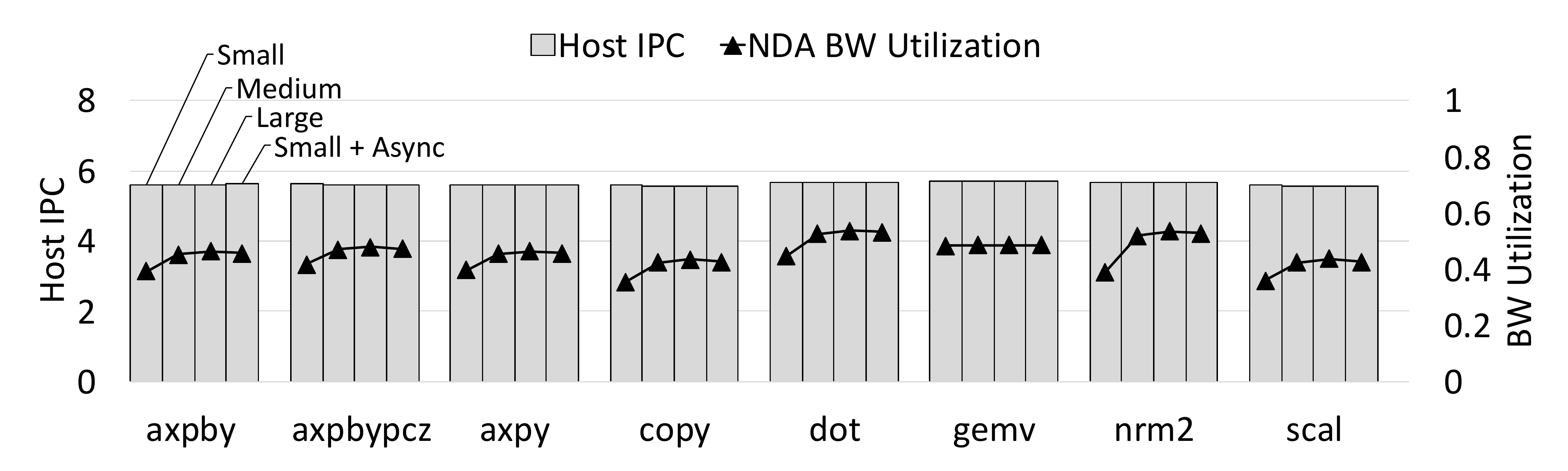}
	\caption{Impact of NDA operations and operand size.}
	\label{fig:eval_nda_workload}
\end{figure}

\medskip
\noindent\textbf{\textit{Impact of Write-Intensity and Input Size.}}
\fig{fig:eval_nda_workload} shows host and NDA performance when different types of NDA operations are executed with different input sizes. The host application mix with the highest memory intensity (mix1) and the next-rank prediction mechanism is used. In addition, to identify the impact of input size, three different vector sizes are used: small (8KB/rank), medium (128KB/rank), and large (8MB/rank). We evaluate asynchronous launches with the small vector size. We evaluate GEMV with three matrix sizes, where the number of columns is equal to each of the three vector sizes and the number of rows fixed at 128.

Overall, performance is inversely related to write intensity, and short execution time per launch results in low NDA performance. The NRM2 operation with the small input has the shortest execution time. Because of its short execution time, NRM2 is highly impacted by the launching overhead and load imbalance caused by concurrent host access. On the other hand, GEMV executes longer than other operations and it is impacted less by load imbalance and launching overhead. With the asynchronous launch optimization, the impact of load imbalance decreases and NDA bandwidth increases.

\medskip
\noindent\fbox{\begin{minipage}{0.46\textwidth}
Takeaway 4: Asynchronous launch mitigates the load imbalance caused by short-duration NDA operations.
\end{minipage}}

\begin{figure}[t!]
\centering
	\includegraphics[width=0.43\textwidth]{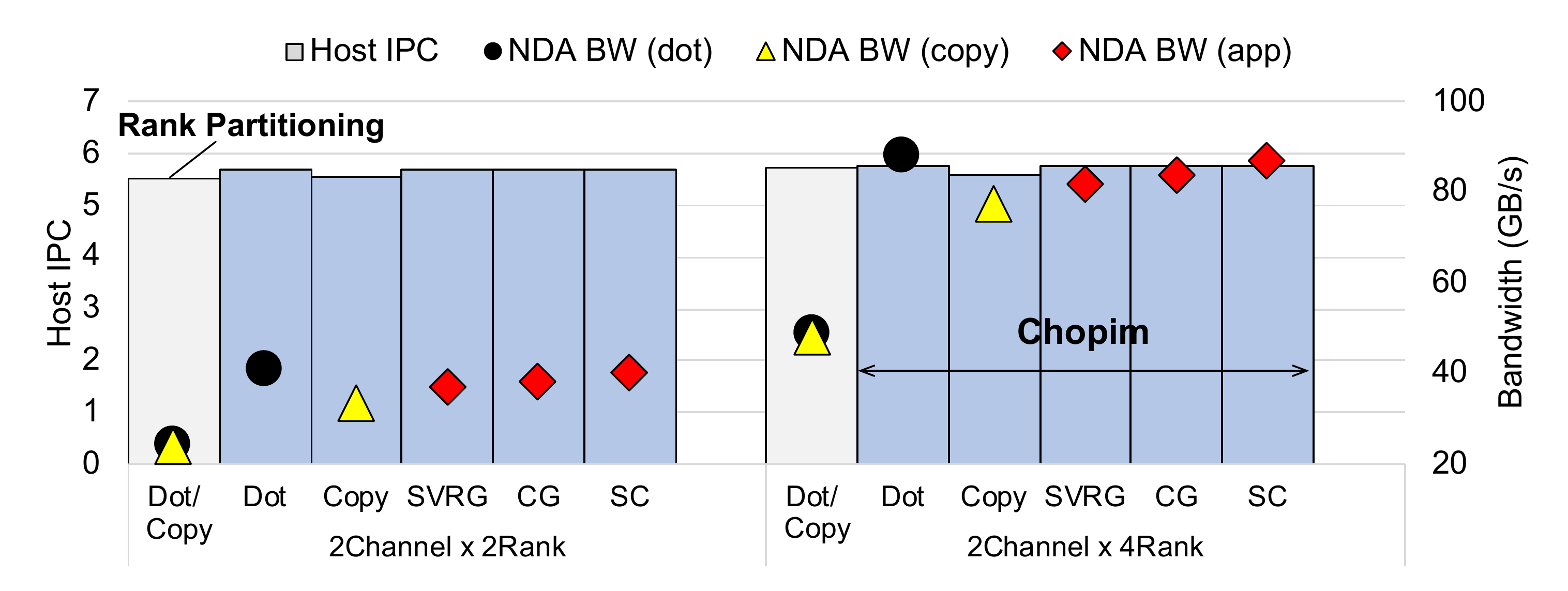}
	\caption{Scalability Chopim vs.~rank partitioning.}
	\label{fig:eval_scal}
	\vspace*{-4mm}
\end{figure}

\medskip
\noindent\textbf{\textit{Scalability Comparison.}}
\fig{fig:eval_scal} compares Chopim with the performance of rank partitioning (RP). For RP, we assume that ranks are evenly partitioned between the host and NDAs. Since read- and write-intensive NDA operations show different trends, we separate those two cases. Other application results (SVRG, CG, and SC) are shown to demonstrate that their performance falls between these two extreme cases.\bcut{We do not evaluate SVRG with RP because it disallows sharing.} We use the most memory-intensive mix1 as the host workload. 
The first cluster shows performance when the baseline DRAM system is used. For both the read- and write-intensive NDA workloads, Chopim performs better than rank partitioning. This shows that opportunistically exploiting idle rank bandwidth can be a better option than dedicating ranks for acceleration. The second cluster shows performance when the number of ranks is doubled. Compared to rank partitioning, Chopim shows better performance scalability. While NDA bandwidth with rank partitioning exactly doubles, Chopim more than doubles due to the increased idle time per rank. SVRG results fall between extreme DOT and COPY cases.  

\medskip
\noindent\fbox{\begin{minipage}{0.46\textwidth}
    Takeaway 5: Chopim scales better than rank partitioning because short issue opportunities grow with rank count.
\end{minipage}}

\begin{figure}[t!]
\centering
	\subfloat [Convergence over time with and without NDA.] {
		\includegraphics[width=0.43\textwidth]{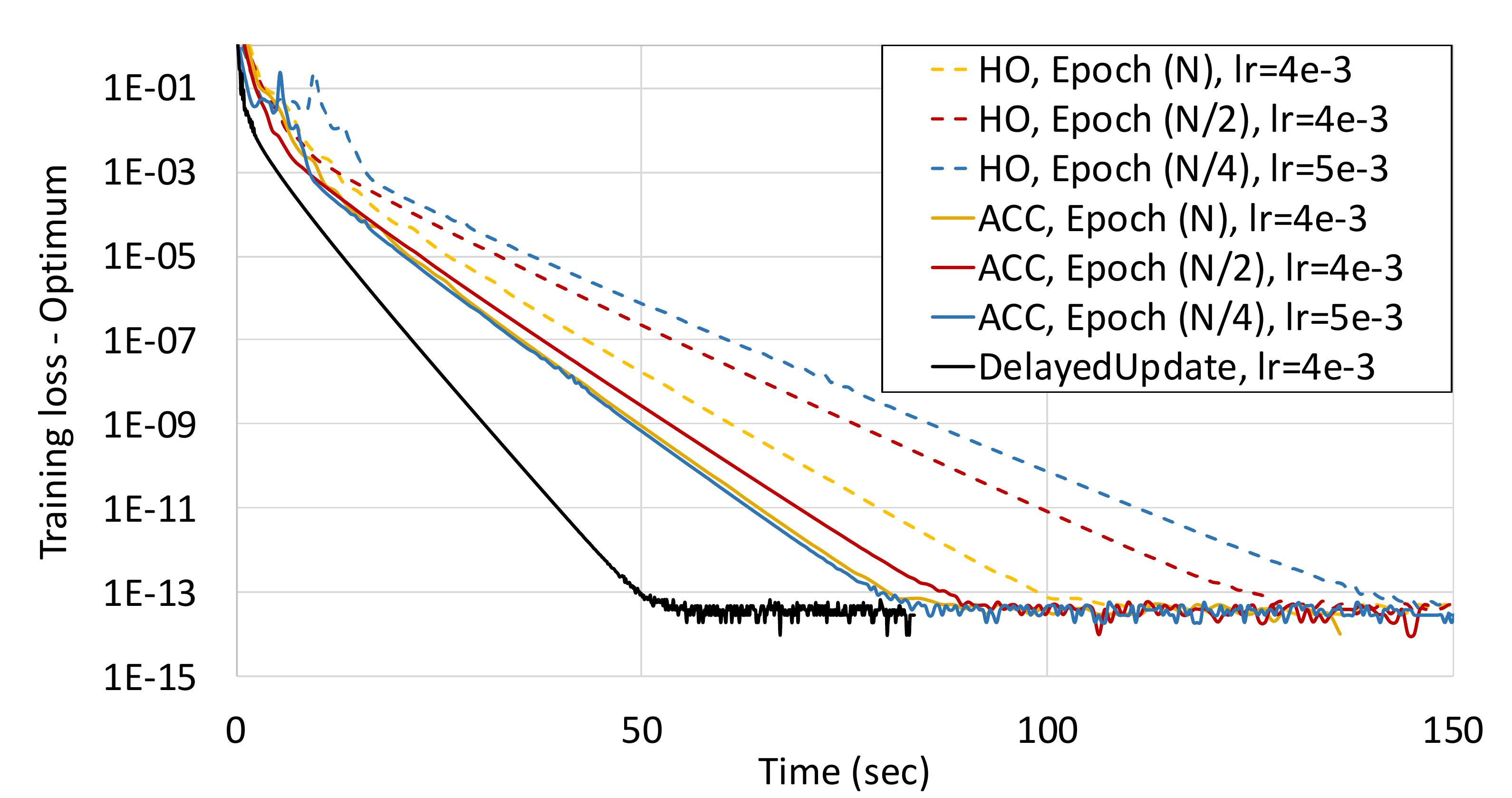}
		\label{fig:eval_svrg}
	} \\
	\subfloat [NDA speedup scaling (normalized to host only).] {
		\includegraphics[width=0.37\textwidth]{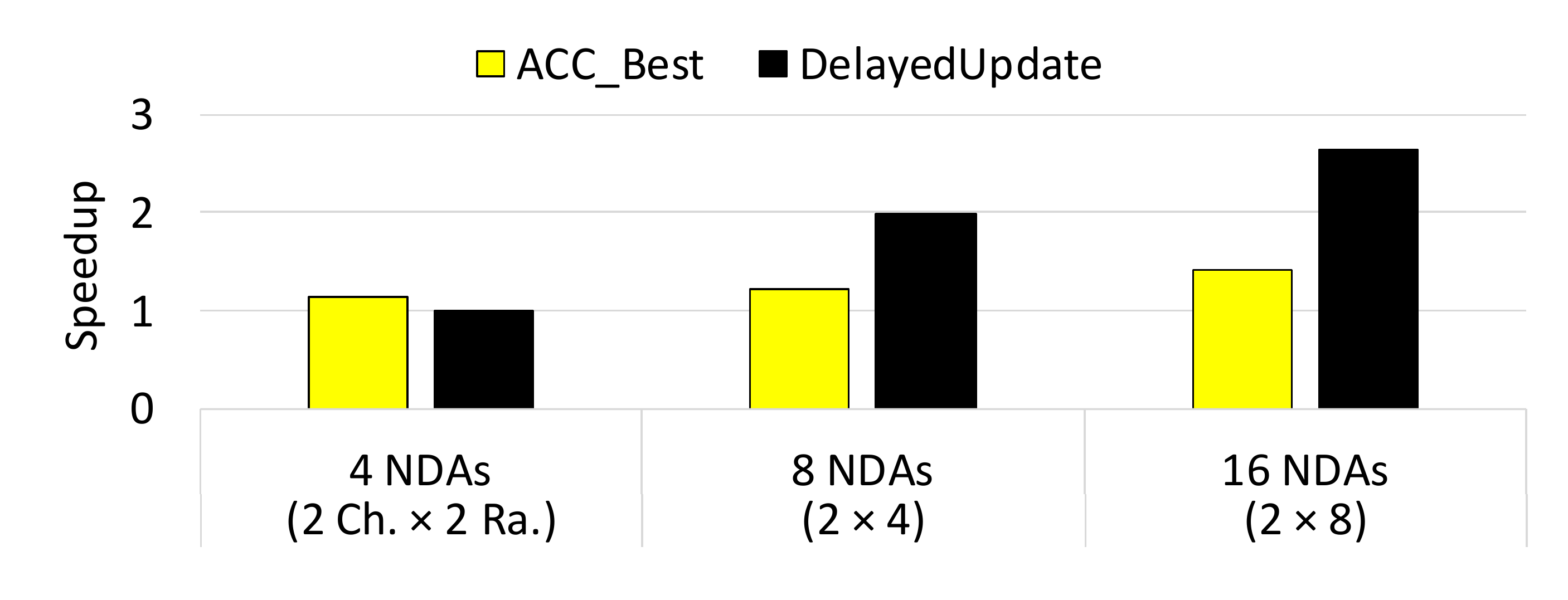}
		\label{fig:eval_svrg_speedup}
	} \\
	\caption{Impact of NDA summarization in SVRG with and without delayed update (HO: Host-Only, ACC: Accelerated with NDAs, ACC\_Best: Best among all ACC options).}
	\label{fig:eval_svrg_results}
	\vspace*{-4mm}
\end{figure}

\medskip
\noindent\textbf{\textit{SVRG Collaboration Benefits.}}
\fig{fig:eval_svrg} shows the convergence results with and without NDA (8 NDAs). We use a shared memory region to enable concurrent access to the same data and the next-rank prediction mechanism is used. Compared to the host-only case, the optimal epoch size decreases from \textit{N} to \textit{N/4} when NDAs are used. This is because the overhead of summarization decreases relative to the host-only case. Furthermore, SVRG with delayed updates gains additional performance demonstrating the benefits made possible by the concurrent host and NDA access when each processes the portion of the workload it is best suited for. Though the delayed update updates the correction term more frequently, the best performing learning rate is lower than ACC with epoch \textit{N/4}, which shows the impact of staleness on the delayed update.

When NDA performance grows by adding NDAs (additional ranks), delayed-update SVRG demonstrates better performance scalability. \fig{fig:eval_svrg_speedup} compares the performance of the best-tuned serialized and delayed-update SVRG with that of host-only with different number of NDAs. We measure performance as the time it takes the training loss to converge (when it reaches $1e-13$ away from optimum). Because more NDAs can calculate the correction term faster, its staleness decreases, consequently, a higher learning rate with faster convergence is possible.

\medskip
\noindent\fbox{\begin{minipage}{0.46\textwidth}
Takeaway 6: Collaborative host-NDA processing on shared data speeds up SVRG logistic regression by 50\%. 
\end{minipage}}

\medskip
\noindent\textbf{\textit{Memory Power.}}
We estimate the power dissipation in the memory system under concurrent access. The theoretical maximum possible power of the memory system is 8W when only the host accesses memory. When the most memory-intensive application mixes are executed, the average power is 3.6W. The maximum power of NDAs is 3.7W and is dissipated when the scratchpad memory is maximally used in the average gradient computation. In total, up to 7.3W of power is dissipated in the memory system, which is lower than the maximum possible with host-only access. This power efficiency of NDAs comes from the low-energy internal memory accesses and because Chopim minimizes overheads.

\medskip
\noindent\fbox{\begin{minipage}{0.46\textwidth}
Takeaway 7: Operating  multiple ranks for concurrent access does not increase memory power significantly. 
\end{minipage}}



%% file: tex/relatedwork.tex

\section{Related Work}
\label{sec:related_work}

To the best of our knowledge, this is the first work that proposes solutions for near data acceleration while enabling the concurrent host and NDA access without data reorganization and in a non-packetized DRAM context. \ykadd{Packetized DRAM, while scalable, may suffer from 2--4x latency longer than DDR-based protocol even under very low or no load \cite{hadidi2018performance}. } To solve this unique problem, many previous studies have influenced our work. 


The study of near data acceleration has been conducted in a wide range as the relative cost of data access becomes more and more expensive compared to the computation itself. The nearest place for computation is in DRAM cells \cite{seshadri2017ambit,li2017drisa,seshadri2015fast} or the crossbar cells with emerging technologies  \cite{li2016pinatubo,chi2016prime,shafiee2016isaac,song2017pipelayer,song2018graphr,sun2017energy,chen2018regan,long2018reram}. Since the benefit of near-data acceleration comes from high bandwidth and low data transfer energy, the benefit becomes larger as computation move closer to memory. However, area and power constraints are significant, restricting adding complex logic. As a result, workloads with simple ALU operations are the main target of these studies. 

3D stacked memory devices enable more complex logic on the logic die and still exploit high internal memory bandwidth. Many recent studies are conducted based on this device to accelerate diverse applications \cite{gao2017tetris,kim2016neurocube,drumond2017mondrian,ahn2016scalable,ahn2015pim,guo20143d,hsieh2016transparent,hsieh2016accelerating,liu2017concurrent,pattnaik2016scheduling,zhang2014top,gao2015practical,nair2015active,hong2016accelerating,boroumand2016lazypim,liu2018processing,boroumand2019conda}. However, in these proposals, the main memory role of the memory devices has gained less attention compared to the acceleration part. Some prior work \cite{akin2015hamlet,sura2015data,akin2016data,boroumand2018google} attempts to support the host and NDA access to the same data but only with data reorganization and in a packetized DRAM context. Parrnaik et al. \cite{pattnaik2016scheduling} show the potential of concurrently running both the host and NDAs on the same memory. However, they assume an idealized memory system in which there is no contention between NDA and host memory requests. We do not assume this ideal case. The main contributions of Chopim are precisely to provide mechanisms for mitigating interference.

On the other hand, \textit{NDA} \cite{farmahini2015nda}, Chameleon \cite{asghari2016chameleon}, and MCN DIMM \cite{alian2018nmp} are based on conventional DIMM devices and changes the DRAM design to practically add PEs.\bcut{\textit{NDA} finds the places to add TSVs from commodity DDR3 devices and solves data layout problem by shuffling. It also proposes solutions to switch mode between host and NDA (precharge-all method) and to avoid concurrent host access (rank partitioning). Chameleon finds an opportunity for near-data acceleration in Load-Reduced DIMM and places PEs in data buffer chips. To overcome the command bandwidth bottleneck, they split DQs and use a part of them for transferring memory commands for PEs. MCN DIMM realized DIMM-type NDAs by enabling the host and MCN processors to communicate with network protocol but via DDR interface. Each MCN DIMM runs a light-weight OS and acts as a small independent computing node. Based on this prior work, we focus more on  host-NDA concurrent access.} Unlike rank partitioning and coarse-grain mode switching used in the prior work, we let host and PEs share ranks to maximize parallelism and partition banks to decrease contention.

%% file: tex/conclusion.tex
\section{Conclusion} 
\label{sec:conclusion}

In this paper, we introduced solutions to share ranks and enable concurrent access between the host and NDAs. Instead of partitioning memory in coarse-grain manner, both temporally and spatially, we interleave accesses in fine-grain manner to leverage the unutilized rank bandwidth. To maximize bandwidth utilization, Chopim enables coordinating state between the memory controllers of the host and NDAs in low overhead, to reduce extra bank conflicts with bank partitioning, to efficiently block NDA write transactions with stochastic issue and next-rank prediction to mitigate the penalty of read/write turnaround time, and to have one data layout that allows the host and NDAs to access the same data and realize high performance. Our case study also shows that collaborative execution between the host and NDAs can provide better performance than using just one of them at a time. Chopim offers insights to practically enable NDA while serving main memory requests in real systems and enables more effective acceleration by eliminating data copies and encouraging tighter host-NDA collaboration.
